\documentclass[showpacs,twocolumn,superscriptaddress,prb,preprintnumbers,nofootinbib]{revtex4}
\usepackage{amsmath,amssymb,bm,graphicx,color}

\begin{document}

% ==============================================================================
%\title{Dynamical properties of the $XXZ$ model on a honeycomb lattice in a field}
\title{Field-induced dynamical properties of the $XXZ$ model on a honeycomb lattice}
% ==============================================================================
\author{P. A. Maksimov}
\thanks{Corresponding author. E-mail: pmaksimo@uci.edu}
\affiliation{Department of Physics and Astronomy, University of California, Irvine, California 92697, USA}
\author{A. L. Chernyshev}
\affiliation{Department of Physics and Astronomy, University of California, Irvine, California 92697, USA}
% ==============================================================================
% ==============================================================================
\date{\today}
% ==============================================================================

% ==============================================================================
\begin{abstract}
We present a comprehensive $1/S$ study of the field-induced dynamical properties of the nearest-neighbor 
$XXZ$ antiferromagnet on a honeycomb lattice using the formalism of the 
nonlinear spin-wave theory developed for this model. The external magnetic field controls spin frustration in the system 
and induces non-collinearity of the spin structure, which is essential for the two-magnon decay processes.
Our results include an intriguing field-evolution of the regions of the Brillouin zone wherein decays of 
spin excitations are prominent, a detailed classification of the decay channels involving magnons from
both excitation branches, and a thorough analysis of the singularities in the magnon spectra 
due to coupling to the two-magnon continuum, all of which are illustrated for several field and 
anisotropy values.  We highlight a number of features related to either the non-Bravais nature 
of the lattice, or the existence of the Dirac-like points in the spectrum. In addition,
the asymptotic behavior of the decay rates near high-symmetry points is analyzed in detail. 
The inelastic neutron-scattering spin-spin structure factor is obtained in the 
leading $1/S$ order and is shown to exhibit qualitatively distinct fingerprints of the decay-induced magnon dynamics
such as quasiparticle peaks broadened by decays and strong spectral weight redistribution. 
\end{abstract}
% ==============================================================================

% ==============================================================================
% ==============================================================================
\maketitle
% ==============================================================================
\section{Introduction}
% ==============================================================================

Quantum spin models on the honeycomb lattice attract significant interest for several interrelated reasons. 
First are the proposals of exotic spin-liquid and valence-bond states \cite{Kitaev,Sheng,Zhu,Fakher,Sid,Flint} together
with a potential experimental realization of some of them in iridium-based compounds and other systems.
\cite{Jackeli,Perkins} Second is the possibility of inducing various forms of magnetism in graphene 
structures.\cite{Uchoa08,grapheneYIG}
Third is due to a number of intriguing  order by disorder effects\cite{Paramekanti,italians} and unusual ordered 
ground states in cases when anisotropy and frustration are involved.
\cite{Galitsky,Zhu1,Sheng1,Bishop,Oitmaa11,Zh13,Kopietz} 

Historically, the honeycomb lattice was also considered as a path to enhanced quantum fluctuations in the 2D spin
models due to the low coordination number of  nearest neighbors.\cite{Singh}
The majority of theoretical studies of these models are devoted to the ground-state problem with the spectral
properties receiving less attention.\cite{Knolle} 
Thus, the main focus of the present work is on the role of potentially stronger effect of quantum fluctuations in 
the spectral properties of the honeycomb-lattice spin systems, the subject relevant to a number of  
experimental systems that have recently become available.
There is a significant  variety of materials
related to this interest, in many of which excitation spectra have been investigated by the inelastic
neutron scattering.\cite{oldLPR,Zaliznyak,oxynitrate,Coldea,usLPR,Matsuda10}
 
In this work we present a consideration of the spectral properties of the nearest-neighbor $XXZ$ model 
on a honeycomb lattice, in which frustration is induced by external magnetic field. 
While the ground state in  this case is a fairly trivial canted modification of the N\'{e}el state, 
it can be expected that the spectral properties may become rather involved due to the field-induced non-collinearity 
of the spin structure and to the concomitant decays of spin excitations.
Such an expectation is based on the studies of the decay-induced dynamical properties of the square-lattice 
antiferromagnets in a field,\cite{99,Mourigal10,Fuhrman13,RMP,Sulyasen,Lauchli,Kreisel,exper} in which drastic 
modifications of the spectra and various forms of singular behavior have been documented.

In the present work, we provide a similar systematic analysis of the decay-induced dynamics for the 
honeycomb-lattice antiferromagnets. Although both the square and the honeycomb lattices are bipartite 
and thus one can expect a close similarity of the results, there are several important differences
that make the honeycomb case significantly richer.
First, the honeycomb lattice is non-Bravais  with two distinct magnon modes present. In an anisotropic case and for
any value of the field, the decays of the ``optical''-like mode into two magnons from the ``acoustic'' branch are
kinematically allowed, while in the square lattice the field must exceed a threshold value for that to happen.\cite{99,RMP} 
Then, for the square lattice, the ${\bf k}\!=\!0$ mode is protected against decays as it is associated with the uniform precession 
of the field-induced magnetization.
One of the interesting questions about the honeycomb case is whether the ${\bf k}\!=\!0$ ``optical'' mode 
has a similar protection or it is allowed to decay.
Second, the honeycomb lattice is not inversion-symmetric with respect to the lattice points. As we
will demonstrate, this feature complicates analytical aspects of the nonlinear spin-wave theory
because of the complex magnon hopping amplitudes. It is also \emph{a priori} 
not obvious what differences can follow from that. In addition, the Dirac-like degeneracy points connecting 
magnon branches are present in the spectra of the $XXZ$ model at the ${\bf K}$-points of the Brillouin zone, 
modifying kinematic consideration for the decays  and 
for the associated singularities. Lastly, due to a larger role 
of  quantum fluctuations in the ground state,\cite{Singh} one can also expect 
a significantly enhanced role of the decays in the spectrum of the honeycomb-lattice models compared
to the square-lattice case.

We thus offer  a comprehensive study of the decay-induced dynamical effects in the
nearest-neighbor  $XXZ$ model on a honeycomb lattice in a field using the $1/S$ approach.
One of the achievements of the present work is the formulation of the nonlinear spin-wave theory 
for the honeycomb-lattice models. While we ignore the contributions of the higher-order 
$1/S$ terms and of the non-decay corrections of the same order, one can still expect a qualitative 
applicability of our results even for the $S\!=\!1/2$ case, similarly to the problems studied previously.\cite{RMP,triangle}  
Extensions of our consideration to the $J_1\!-\!J_2$ model as well as to the other cases can also 
be anticipated to retain main qualitative features of our findings. 
One can anticipate an immediate applicability of our results to realistic materials discussed 
in Refs.~\onlinecite{Zaliznyak,Matsuda10,oxynitrate}, which exhibit ordered ground states. We expect 
that applying magnetic field will result in a redistribution of 
spectral weight according to the predictions of our work. 
Next-nearest-neighbor interactions\cite{Zaliznyak} should not affect the results significantly.

The paper is organized as follows. In Sec.~\ref{sec_swt} we present the model Hamiltonian and develop 
the nonlinear  spin-wave formalism. In Sec.~\ref{sec_kinematics} we analyze  kinematic conditions that define 
decay regions and singularity contours in the magnon spectra and study their field evolution.
In Sec.~\ref{sec_cubic}, damping is calculated in the Born approximation  for both magnon branches in different 
decay channels and throughout the Brillouin zone for several representative values of anisotropy and   field. 
Here we also analyze the origin of singularities in  decays
and present asymptotic behaviors of the decay rate in the vicinities of the high-symmetry points. 
In Sec.~\ref{sec_strfac}, the spin-spin structure factor calculations are presented and several characteristic features
of its behavior are discussed.

% ==============================================================================
\section{Nonlinear spin-wave theory}
% ==============================================================================
\label{sec_swt}
% ==============================================================================

% ==============================================================================
\begin{figure}[b]
\includegraphics[width=0.999\linewidth]{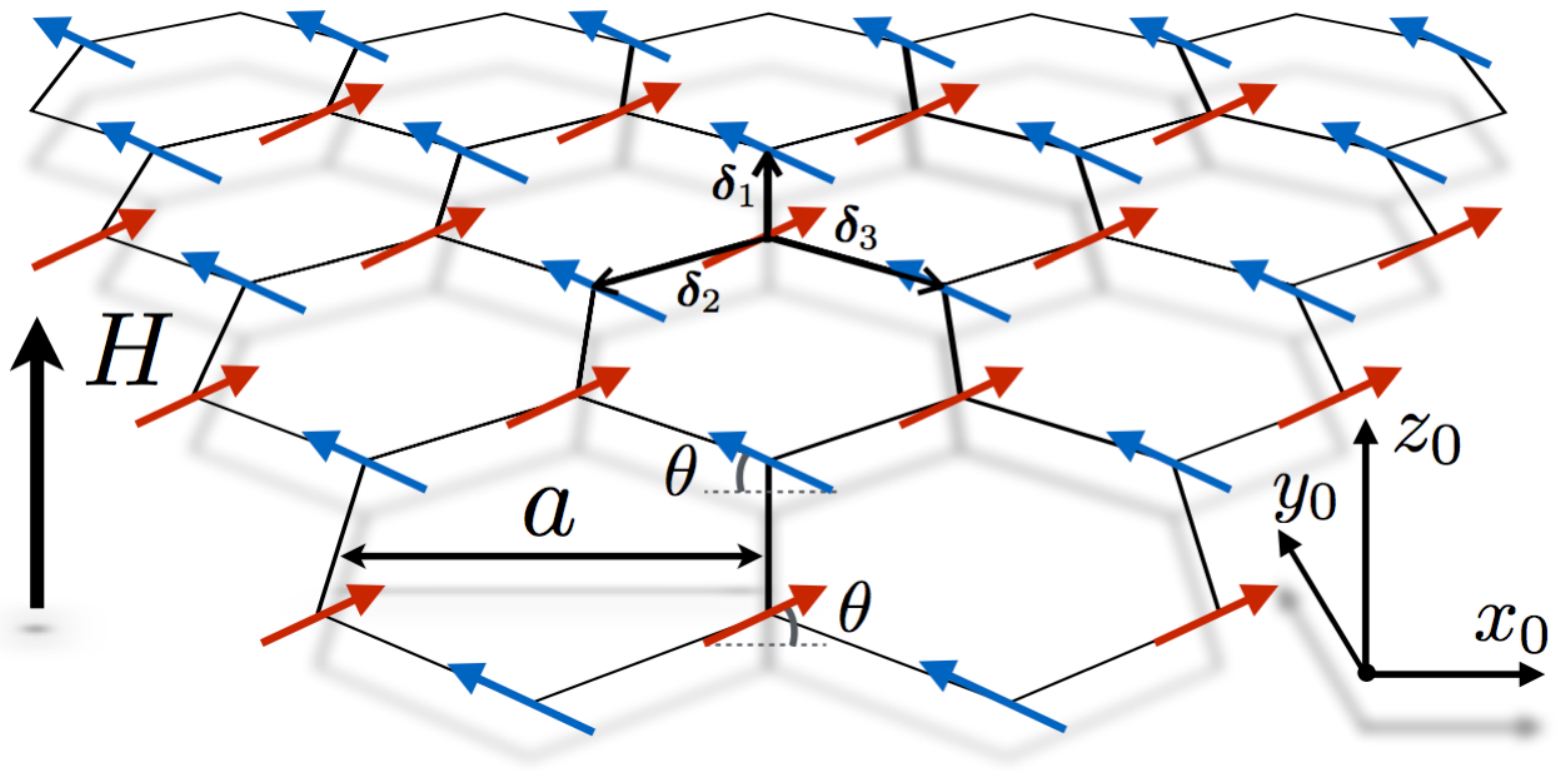}
\caption{(Color online) A sketch of the canted antiferromagnetic 
spin configuration on the honeycomb lattice. The nearest-neighbor vectors ${\bm \delta}_i$, lattice constant $a$, 
canting angle $\theta$,
laboratory reference frame $\{x_0,y_0,z_0\}$, and the field direction along the laboratory axis $z_0$ are indicated.}
\label{fig_lattice}
\end{figure}
% ==============================================================================

In this Section, we provide an exposition of the $1/S$ approach to the nearest-neighbor honeycomb-lattice
$XXZ$ antiferromagnet in a field.  While the linear spin-wave theory (LSWT) treatment of the $H\!=\!0$ case
is well-known,\cite{Singh,Kopietz,Paramekanti} the role of the nonlinear, anharmonic terms for this model,
to the best of our knowledge, has not been examined previously. Since the corresponding derivation also includes 
several steps that are unfamiliar from the previously studied case of the square-lattice model in a field,
\cite{RMP} we present it here in detail. We would also like to note that the structure of the derived 
anharmonic terms bears a close similarity to the ones in the other models on the non-Bravais lattices, 
such as kagome-lattice model studied recently,\cite{kagome14,kagome15} and we attempt to preserve
a generality in the respective notations.

We begin with the nearest-neighbor $XXZ$ Hamiltonian on a honeycomb lattice in external field, which is  
applied parallel to the $z_0$-axis, at $T=0$
\begin{eqnarray}
\hat{\cal H}=J\sum_{\langle ij\rangle}\Big({\bf S}_i \cdot {\bf S}_j-\left(1-\Delta\right) S^{z_0}_i S^{z_0}_j\Big)
-H\sum_{i} S^{z_0}_i\ ,
\label{eq_hamiltonian1}
\end{eqnarray}
where  the sum is over the nearest-neighbor bonds $\langle ij\rangle$, $J$ is  an exchange coupling constant, 
$0\!\leq\Delta\!\leq\!1$ is the easy-plane anisotropy parameter, and $H$ is an external magnetic field in units of $g \mu_B$. 

% ==============================================================================
\subsection{Spin transformation}
% ==============================================================================

In zero magnetic field spins align in a classical N\'{e}el structure. In an applied field,  one can expect the 
spins to cant towards the field direction. Because of that, we need to align the local spin quantization axis 
on each site in the direction given by the canted spin configuration, see Fig.~\ref{fig_lattice},
with the canting angle to be defined from the energy minimization. 
The corresponding general transformation of the spin components from the laboratory reference frame 
$\{x_0,y_0,z_0\}$ to the local reference frame $\{x,y,z\}$ can be performed using two consequent rotations
\begin{eqnarray}
{\bf S}^0_i={\bf R}_{{\bf Q},\beta_i}\cdot {\bf R}_{\theta} \cdot{\bf S}_i\, ,
\label{eq_frametransform}
\end{eqnarray}
where the matrix ${\bf R}_{{\bf Q},\beta_i}$ corresponds to the rotations in the  $x_0y_0$ plane
\begin{eqnarray}
{\bf R}_{{\bf Q},\beta_i}=
\left( \begin{array}{ccc} 
\cos \varphi_i & -\sin\varphi_i & 0\\ 
\sin\varphi_i & \cos\varphi_i & 0\\
0 & 0 & 1 
\end{array}\right),
\label{Ri}
\end{eqnarray}
where $\varphi_i={\bf Q} \cdot {\bf r}_i+\beta_i$ with
$\textbf{Q}$ being the ordering wavevector of the spin configuration 
and $\beta_i$ the phase shift inside the unit cell. 
The matrix ${\bf R}_{\theta}$ performs  spin rotation within the $x_0z_0$ towards the field
\begin{eqnarray}
{\bf R}_{\theta}=\left( \begin{array}{ccc} \sin \theta& 0 & \cos \theta \\ 
0 & 1 & 0\\
-\cos\theta & 0 & \sin\theta \end{array}\right),
\label{Rth}
\end{eqnarray}
where $\theta$ is the canting angle, see Fig.~\ref{fig_lattice}.

Since, by construction, all spins are oriented along their local $z$ axes, 
the classical energy can be obtained from (\ref{eq_hamiltonian1}) using transformations (\ref{Ri}) and (\ref{Rth})
\begin{eqnarray}
\frac{E_{\rm cl}}{NJS^2}=
\sum_{{\bm\delta}_i}\Big(\cos^2\theta\cos \delta\varphi_i+\Delta\sin^2 \theta\Big)
-\frac{2H\sin\theta}{JS}\, ,
\label{Ecl}
\end{eqnarray}
where $N$ is the number of unit cells,
$\delta\varphi_i\!=\!{\bf Q}\cdot {\bm\delta}_i\!+\!\beta$ 
with $\beta \!=\!\beta_i\!-\!\beta_j$, and ${\bm\delta}_i$ are the nearest-neighbor vectors
\begin{eqnarray}
{\bm\delta}_1=\left( 0,\frac{a}{\sqrt{3}}\right),~{\bm\delta}_2=-\left(\frac{a}{2},\frac{a}{2\sqrt{3}} \right),
~{\bm\delta}_3=-{\bm\delta}_1-{\bm\delta}_2,
\end{eqnarray}
where $a$ is the lattice constant. 

Energy minimization in (\ref{Ecl}) yields a unique solution
\begin{eqnarray}
{\bf Q}=0,~\beta =\beta_i-\beta_j=\pi,~ \sin\theta=H/H_s,
\label{eq_classic_parameters}
\end{eqnarray}
where $H_s=3JS(1+\Delta)$. As is expected, the spins form a canted N\'{e}el structure shown in Fig.~\ref{fig_lattice}. 
With these parameters the general transformation in (\ref{eq_frametransform}) becomes
\begin{eqnarray}
&&S^{x_0}_{A(B)}=\pm \sin\theta\, S^x_{A(B)}\pm\cos\theta \, S^z_{A(B)},\nonumber\\
&&
\label{eq_localtransform}
S^{y_0}_{A(B)}=\pm S^y_{A(B)},\\
&&S^{z_0}_{A(B)}=-\cos\theta \, S^x_{A(B)}+\sin\theta\, S^z_{A(B)}, \nonumber
\end{eqnarray}
where $A$ and $B$ are the two sublattices of the structure.

With that, one can rewrite the Hamiltonian in (\ref{eq_hamiltonian1}) in the local reference frames of spins
as a sum of two parts
\begin{eqnarray}
\hat{\cal H}=\hat{\cal H}_{\rm even}+\hat{\cal H}_{\rm odd}\, ,
\label{Heven_odd}
\end{eqnarray}
where the classification in even/odd is done in anticipation of the spin bosonization, which is performed next.
The even part is given by
\begin{eqnarray}
\label{Heven}
&&\hat{\cal H}_{\rm even}=J\sum_{\langle ij\rangle}\Big( 
\left[\Delta\cos^2\theta-\sin^2\theta\right]S^x_i S^x_{j}-S^y_i S^y_{j}\quad\quad \\
&&\phantom{\hat{\cal H}_{\rm even}}
+\left[\Delta\sin^2\theta-\cos^2\theta\right]S^z_i S^z_{j}\Big)-H\sin\theta\sum_i  S^z_i, \nonumber
\end{eqnarray}
and the odd terms are
\begin{eqnarray}
&&\hat{\cal H}_{\rm odd}=-\frac12 \, J\sin 2\theta\, 
(1+\Delta) \sum_{\langle ij\rangle}\Big(S^z_i S^x_{j}+S^x_i S^z_{j}\Big)\nonumber\\
&&\phantom{\hat{\cal H}_{\rm odd}} +H \cos\theta \sum_i S^x_i.
\label{eq_Hodd}
\end{eqnarray}
The subsequent treatment of the spin Hamiltonian involves a standard Holstein-Primakoff transformation 
on each site,\cite{RMP} bosonizing spin operators via
\begin{equation}
S^{+}_i=a_i\sqrt{2S-a^\dagger_i a_i}, \quad S^z_i=S-a^\dagger_i a_i \,
\label{HP}
\end{equation}
with the subsequent expansion of the square roots  in $\langle a^{\dagger} a\rangle /2S$, 
yielding the series 
\begin{eqnarray}
\hat{\cal H}=\hat{\cal H}^{(0)}+\hat{\cal H}^{(1)}+\hat{\cal H}^{(2)}+\hat{\cal H}^{(3)}+\dots ,
\label{eq_expansion}
\end{eqnarray}
where the numbers in the superscript correspond to the number of bosonic operators in the given term 
and $\hat{\cal H}^{(n)}\!\propto\!S^{2-n/2}$.
 The first term  in (\ref{eq_expansion}) is the classical energy 
 $\hat{\cal H}^{(0)}\!=\!E_{\rm cl}\!=\!-3NJS^2 \left[1+ (\Delta+1) \sin^2 \theta\right]$,
and $\hat{\cal H}^{(1)}$ vanishes at the energy minimum, as usual.

% ==============================================================================
\subsection{Linear spin-wave theory}
% ==============================================================================

The first non-vanishing term in the expansion (\ref{eq_expansion}) beyond the classical energy 
is the quadratic Hamiltonian, which is given by
\begin{eqnarray}
&&\hat{\cal H}^{(2)}=JS  \sum_{\langle ij\rangle}\Big[a^\dagger_{1i} a^{\phantom{\dag}}_{1i} 
+a^\dagger_{2j} a^{\phantom{\dag}}_{2j}
-\lambda\left(a^\dagger_{1i} a^{\phantom{\dag}}_{2j}+a^\dagger_{2i} a^{\phantom{\dag}}_{1j}\right)
\nonumber\\
&&\phantom{\hat{\cal H}^{(2)}=JS  \sum_{\langle ij\rangle}}
+(1-\lambda)\left(a^{\phantom{\dag}}_{1i} a^{\phantom{\dag}}_{2j} +a^\dagger_{1i} a^\dagger_{2j}\right)\Big],
\label{H2}
\end{eqnarray}
where $a^{(\dag)}_{1(2)}$ are the operators on the sublattices $A(B)$. 
It is worth noting that, for this Hamiltonian, field and anisotropy can be combined in a single parameter 
$\lambda$
\begin{eqnarray}
\lambda=1-\frac{1+\Delta}{2}\cos^2\theta\,\quad \mbox{or} \ \ \ 
\frac{H}{H_s}=\sqrt{\frac{\Delta+2\lambda-1}{1+\Delta}}. \ \ 
\label{lambda}
\end{eqnarray} 
At the saturation, $H\!=\!H_s$, $\lambda\!=\!1$ for any $\Delta$ and at smaller fields, 
$0\!\leq\!H\!\leq\!H_s$, $\lambda$ is within the range $(1-\Delta)/2\!\leq\!\lambda\!\leq\!1$.

Next, we introduce  Fourier transformation
\begin{eqnarray}
a^{\phantom{\dag}}_{\alpha i}=
\frac{1}{\sqrt{N}} \sum_{{\bf k}} e^{i{\bf k} \left( {\bf r}_i+{\bm\rho}_\alpha\right)} 
a^{\phantom{\dag}}_{\alpha {\bf k}}\, ,
\label{eq_fourier}
\end{eqnarray}
where ${\bf r}_i$ are the coordinates of the unit cell and 
${\bm\rho}_\alpha$ are coordination vectors inside the unit cell for $\alpha\!=\!1,2$ atoms, 
${\bm\rho}_1\!=\! (0,0)$ and ${\bm\rho}_2\!=\! {\bm \delta}_1\!=\!(0,a/\sqrt{3})$. 
This gives
\begin{eqnarray}
&&\hat{\cal H}^{(2)}=3JS\sum_{{\bf k}} 
\Big[a^\dagger_{1{\bf k}} a^{\phantom{\dag}}_{1{\bf k}} 
+a^\dagger_{2{\bf k}} a^{\phantom{\dag}}_{2{\bf k}}
\label{H2k}\\
&&\phantom{\hat{\cal H}^{(2)}}
-\left(\lambda\gamma_{{\bf k}}a^\dagger_{2{\bf k}} a^{\phantom{\dag}}_{1{\bf k}}
-(1-\lambda)\gamma_{{\bf k}}a^{\phantom{\dag}}_{1{\bf k}} a^{\phantom{\dag}}_{2-{\bf k}} 
+{\rm H.c.}\right)\Big], \nonumber
\end{eqnarray}
where $\gamma_{{\bf k}}$ is the complex nearest-neighbor  amplitude 
\begin{eqnarray}
\gamma_{{\bf k}}=\frac{1}{3}\sum_{{\bm \delta}_i} e^{i{{\bf k}}\cdot {\bm \delta}_i}=
|\gamma_{{\bf k}}|e^{i\varphi_{{\bf k}}},
\label{gk}
\end{eqnarray}
with the phase, which is antisymmetric with respect to ${\bf k}\!\rightarrow\!-{\bf k}$,
$\varphi_{-{\bf k}}=-\varphi_{\bf k}$,
and its absolute value is given by
\begin{eqnarray}
|\gamma_{{\bf k}}|=\frac13\,\sqrt{1+4\cos^2 \tilde{k}_x+4\cos \tilde{k}_x
\cos\tilde{k}_y}\, ,
\end{eqnarray}
where $\tilde{k}_x\!=\!k_x a/2$ and $\tilde{k}_y\!=\!k_y a\sqrt{3}/2$.

The diagonalization of $\hat{\cal H}^{(2)}$ in (\ref{H2k}) proceeds in two steps.\cite{Kopietz,kagome14}
First is a unitary transformation from $a^{\phantom{\dag}}_{1(2){\bf k}}$  to 
a set of their symmetric (antisymmetric) combinations
\begin{eqnarray}
a^{\phantom{\dag}}_{\alpha{\bf k}}=
\frac{e^{i(-1)^\alpha \varphi_{\bf k}/2}}{\sqrt{2}}\sum_{\mu} V^{\alpha\mu}  c^{\phantom{\dag}}_{\mu{\bf k}}\,  ,
\label{eq_phaseshift}
\end{eqnarray}
where the $2\times 2$ matrix ${\bf \hat{V}}$ is 
\begin{eqnarray}
{\bf \hat{V}}=\left( \begin{array}{ccc} 1& 1 \\ 1 & -1 \end{array}\right).
\label{V}
\end{eqnarray}
We note that our phase convention in this transformation 
differs from the previous works,\cite{Singh,Kopietz,Paramekanti}
where the phase factor from $\gamma_{\bf k}$ is absorbed in the operators of one of the boson species, while
in our case the phase factor is split symmetrically between both species.
This difference has no projections on the linear spin-wave results, but  will be important for 
a highly symmetric structure of the nonlinear terms discussed in the following.

% ==============================================================================
\begin{figure}
\includegraphics[width=0.999\linewidth]{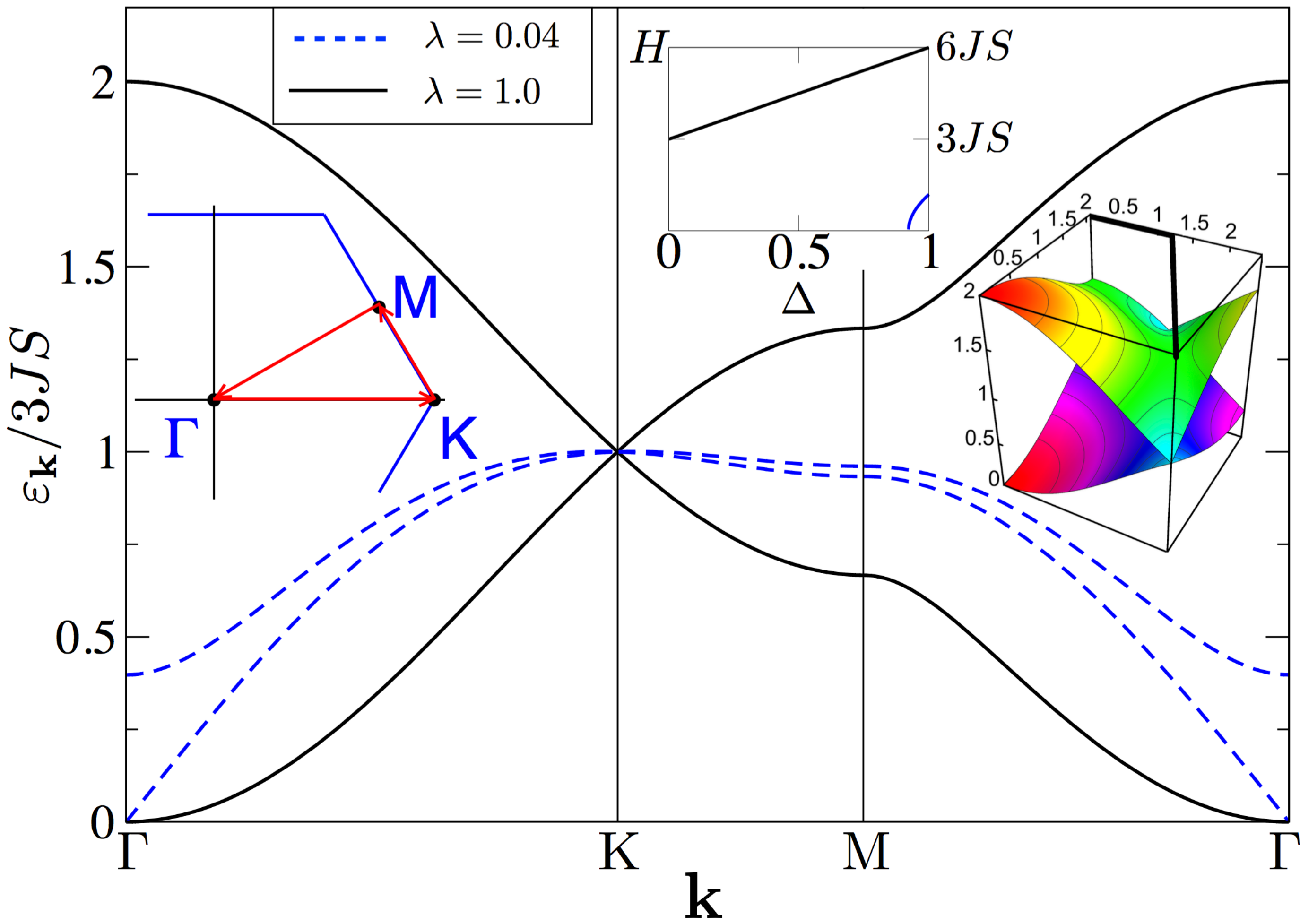}
\caption{(Color online)\ The magnon energies, $\varepsilon_{\mu{\bf k}}$, for two branches 
(\ref{eq_energyspectrum}) along the 
$\Gamma\!\rightarrow\!{\rm K}\!\rightarrow\!{\rm M}\!\rightarrow\! \Gamma$ path (left inset) for two values 
of $\lambda$. Dashed lines are for $\lambda\!=\!0.04$, solid lines for $\lambda\!=\!1$. 
Upper inset shows $H$ vs $\Delta$ that correspond to these values of $\lambda$ via (\ref{lambda}). 
A 3D picture of the energies for $\lambda\!=\!1$ is shown for reference (right inset). High-symmetry
points are ${\rm K}=\left( 4\pi/3a,0 \right)$ and ${\rm M}=\left(\pi/a,\pi/\sqrt{3}a\right)$.}
\label{fig_spectrum}
\vskip -0.3cm
\end{figure}
% ==============================================================================

After this transformation the LSWT Hamiltonian is block-diagonal in the new bosonic index $\mu=1,2$
\begin{equation}
\hat{\cal H}^{(2)}=3JS\sum_{{\bf k},\mu} A_{\mu{{\bf k}}} c^\dagger_{\mu{\bf k}}c^{\phantom{\dag}}_{\mu{\bf k}}
-\frac{B_{\mu{\bf k}}}{2}
\left(c^{\phantom{\dag}}_{\mu{\bf k}}c^{\phantom{\dag}}_{\mu-{\bf k}}+{\rm H.c.}\right),
\end{equation}
where $A_{\mu{\bf k}}$ and $B_{\mu{\bf k}}$ are purely real and are given by
\begin{eqnarray}
A_{\mu{\bf k}}=1+(-1)^\mu\lambda|\gamma_{\bf k}|,~B_{\mu{\bf k}}=(-1)^\mu (1-\lambda)|\gamma_{\bf k}|\, .
\end{eqnarray}
The second step of the diagonalization is a standard Bogolyubov transformation for individual bosonic species
\begin{eqnarray}
c_{\mu{\bf k}}=u_{\mu{\bf k}} d^{\phantom{\dag}}_{\mu{\bf k}}+v_{\mu{\bf k}} d^{\dagger}_{\mu-{\bf k}}\, ,
\label{eq_bogolyubov}
\end{eqnarray}
with the parameters of the transformation given by
\begin{eqnarray}
2u_{\mu{\bf k}}v_{\mu{\bf k}}=\frac{B_{\mu{\bf k}}}{\omega_{\mu{\bf k}}},\quad
u^2_{\mu{\bf k}}+v^2_{\mu{\bf k}}=\frac{A_{\mu{\bf k}}}{\omega_{\mu{\bf k}}},
\label{eq_bogolyubovcoefficients}
\end{eqnarray}
where
%\begin{eqnarray}
$\omega_{\mu{\bf k}}=\sqrt{A^2_{\mu{\bf k}}-B^2_{\mu{\bf k}}}$. %\, .
%\end{eqnarray}
Finally, the excitation spectrum consists of two branches, $\varepsilon_{\mu{\bf k}}\!=\!3JS\omega_{\mu{\bf k}}$, 
which will be referred to as the acoustic and optical modes, see Fig.~\ref{fig_spectrum},
\begin{eqnarray}
\varepsilon_{1(2){\bf k}}=3JS\sqrt{\left(1\mp |\gamma_{\bf k}|\right)\left(1\pm (1-2\lambda)|\gamma_{\bf k}|\right)}\, .
\label{eq_energyspectrum}
\end{eqnarray}
Importantly, 
for any value of anisotropy $\Delta\!<\!1$ and/or for any  
field $H\!>\!0$, $\varepsilon_{2{\bf k}}\!>\!\varepsilon_{1{\bf k}}$ 
in the entire Brillouin zone except for the  ${\rm K}$ and ${\rm K'}$ points, which 
correspond to the Dirac-like degeneracy points in the magnon spectrum. 
At these points, magnon energies are $3JS$ independently of $\Delta$ and $H$, see Fig.~\ref{fig_spectrum}.

% ==============================================================================
\subsection{Nonlinear spin-wave theory}
% ==============================================================================

We note that the two-step diagonalization procedure outlined above, involving subsequent unitary 
and para-unitary transformations, is closely reminiscent of the spin-wave 
approach to the kagome-lattice antiferromagnets\cite{kagome14,kagome15}   and, in principle, 
should be applicable to the other non-Bravais lattices. The following treatment of the anharmonic terms in the 
Hamiltonian expansion in (\ref{eq_expansion}) is also similar, but requires additional care due to complex 
hopping amplitude (\ref{gk}) and to the phase factors in the unitary transformation  (\ref{eq_phaseshift}).

Since we are interested in the decay-induced effects in the spectrum, we need to transform
cubic anharmonic terms from the odd part of the Hamiltonian in  \eqref{eq_Hodd}.
After Holstein-Primakoff transformation (\ref{HP}) it gives 
\begin{eqnarray}
\hat{\cal H}^{(3)}=\frac{J^{(3)}_{\theta,\Delta}}{3}\sum_{\langle ij\rangle} 
\left( a^{\dagger}_{1i} a^{\dagger}_{2j} a^{\phantom{\dag}}_{2j}
+ a^{\dagger}_{2j} a^{\dagger}_{1i} a^{\phantom{\dag}}_{1i} +{\rm H.c.}\right),
\label{H3ij}
\end{eqnarray}
where we have introduced a shorthand notation
\begin{eqnarray}
J^{(3)}_{\theta,\Delta}=\frac{3}{2}\sqrt{\frac{S}{2}}\, J\sin 2\theta\, (1+\Delta) \, .
\label{C3}
\end{eqnarray}
Fourier transformation~\eqref{eq_fourier} of the cubic term (\ref{H3ij}) yields
\begin{eqnarray}
\hat{\cal H}^{(3)}=\sum_{-{\bf p}={\bf k}+{\bf q}}\sum_{\alpha \beta} \left(
G^{\alpha\beta}_{{\bf q}} a^{\dagger}_{\beta{\bf q}} a^{\dagger}_{\alpha{\bf k}} 
a^{\phantom{\dag}}_{\alpha-{\bf p}}+ {\rm H.c.}\right),
\label{H3k}
\end{eqnarray}
where $G^{\alpha\beta}_{{\bf q}}=J^{(3)}_{\theta,\Delta}\,\widetilde{G}^{\alpha\beta}_{{\bf q}}$ and the 
off-diagonal elements, $\alpha\!\neq\!\beta$, of the dimensionless  tensor  are  given by
\begin{eqnarray}
\widetilde{G}^{\alpha\beta}_{{\bf q}}=|\gamma_{\bf q}|e^{i(-1)^{\beta} \varphi_{{\bf q}}}\, , 
\end{eqnarray}
with the diagonal elements  $\widetilde{G}^{\alpha\alpha}_{{\bf q}}\!=\!0$.

The unitary transformation \eqref{eq_phaseshift}  transforms (\ref{H3k}) to
\begin{eqnarray}
\hat{\cal H}^{(3)}=\sum_{-{\bf p}={\bf k}+{\bf q}}\sum_{\eta\nu\mu} \left(
F^{\eta\nu\mu}_{{\bf q},{\bf k}{\bf p}} c^{\dagger}_{\eta{\bf q}} c^{\dagger}_{\nu{\bf k}} 
c^{\phantom{\dag}}_{\mu-{\bf p}}+ {\rm H.c.}\right),
\label{H3ktr}
\end{eqnarray}
where $F^{\eta\nu\mu}_{{\bf q},{\bf k}{\bf p}}\!=\!J^{(3)}_{\theta,\Delta}\,\widetilde{F} ^{\eta\nu\mu}_{{\bf q},{\bf k}{\bf p}}$
with the dimensionless vertex
\begin{eqnarray}
\widetilde{F}^{\eta\nu\mu}_{{\bf q},{\bf k}{\bf p}}=\frac{|\gamma_{\bf q}|}{2\sqrt{2}}\sum_{\alpha\neq\beta}
 V^{\alpha\mu}V^{\alpha\nu}V^{\beta\eta}
 \, e^{i(-1)^\beta \widetilde{\varphi}_{\bf q k p}},
\label{eq_Fphase}
\end{eqnarray}
where the total phase factor 
\begin{eqnarray}
%\Psi_{\bf q k p}=\left( \varphi_{{\bf q}}+ \varphi_{{\bf k}}+ \varphi_{{\bf p}}\right)/2,
\widetilde{\varphi}_{\bf q k p}=\frac{\varphi_{{\bf q}}+ \varphi_{{\bf k}}+ \varphi_{{\bf p}}}{2},
\label{psikqp}
\end{eqnarray}
is introduced for brevity.
The vertex in (\ref{H3ktr}) has an obvious symmetry with respect to permutations of two 
momenta together with the boson indices 
\begin{eqnarray}
\widetilde{F}^{\eta\nu\mu}_{{\bf q},{\bf k}{\bf p}}=\widetilde{F}^{\eta\mu\nu}_{{\bf q},{\bf p}{\bf k}}.
\end{eqnarray}
Using explicit expression for $\widetilde{F}^{\eta\nu\mu}_{{\bf q},{\bf k}{\bf p}}$ in (\ref{eq_Fphase}) 
and for the  unitary transformation in (\ref{eq_phaseshift}), one can considerably simplify 
individual terms of the tensor to
\begin{eqnarray}
&&\widetilde{F}^{111}_{{\bf q},{\bf k}{\bf p}}=\widetilde{F}^{122}_{{\bf q},{\bf k}{\bf p}}=
- \widetilde{F}^{221}_{{\bf q},{\bf k}{\bf p}}=\frac{|\gamma_{\bf q}|}{\sqrt{2}}\cos\widetilde{\varphi}_{\bf q k p},\nonumber\\
&&\widetilde{F}^{112}_{{\bf q},{\bf k}{\bf p}}=\widetilde{F}^{211}_{{\bf q},{\bf k}{\bf p}}=
-\widetilde{F}^{222}_{{\bf q},{\bf k}{\bf p}}=i\frac{|\gamma_{\bf q}|}{\sqrt{2}}\sin\widetilde{\varphi}_{\bf q k p}.
\label{Fcompact}
\end{eqnarray}
Finally, the Bogolyubov transformation (\ref{eq_bogolyubov}) yields the cubic Hamiltonian for the 
magnon normal modes in the following form
\begin{eqnarray}
&&\hat{\cal H}^{(3)}=\frac{1}{3!}\sum_{-{\bf p}={\bf k}+{\bf q}}\sum_{\eta\nu\mu} \left(
\Xi^{\eta\nu\mu}_{{\bf q}{\bf k}{\bf p}} 
d^{\dagger}_{\eta{\bf q}} d^{\dagger}_{\nu{\bf k}} d^{\dagger}_{\mu{\bf p}}+{\rm H.c.}\right)\hskip 0.88cm \
\label{Hsource}
\\
&&\phantom{\hat{\cal H}^{(3)}}
+\frac{1}{2!}\sum_{-{\bf p}={\bf k}+{\bf q}}\sum_{\eta\nu\mu} \left(
\Phi^{\eta\nu\mu}_{{\bf q}{\bf k};{\bf p}} 
d^{\dagger}_{\eta{\bf q}} d^{\dagger}_{\nu{\bf k}} d^{\phantom{\dag}}_{\mu-{\bf p}}+{\rm H.c.}\right),
\label{Hdecay}
\end{eqnarray}
where the combinatorial factors are due to symmetrization in the source (\ref{Hsource}) and decay (\ref{Hdecay}) vertices
\begin{eqnarray}
\Xi^{\eta\nu\mu}_{{\bf q}{\bf k}{\bf p}}=J^{(3)}_{\theta,\Delta}\, \widetilde{\Xi}^{\eta\nu\mu}_{{\bf q}{\bf k}{\bf p}}\,, \quad
\Phi^{\eta\nu\mu}_{{\bf q}{\bf k};{\bf p}}=J^{(3)}_{\theta,\Delta}\, \widetilde{\Phi}^{\eta\nu\mu}_{{\bf q}{\bf k};{\bf p}}
\end{eqnarray}
with the corresponding dimensionless vertices given by 
\begin{eqnarray}
&&\widetilde{\Xi}^{\eta\nu\mu}_{{\bf q}{\bf k}{\bf p}} = 
\widetilde{F}^{\eta\nu\mu}_{{\bf q},{\bf k}{\bf p}} 
(u_{\eta{\bf q}}+v_{\eta{\bf q}})
(u_{\nu{\bf k}}v_{\mu{\bf p}}+v_{\nu{\bf k}}u_{\mu{\bf p}})
\label{eq_sourcevertex}
\\
&&\phantom{\widetilde{\Xi}^{\eta\nu\mu}_{{\bf q}{\bf k}{\bf p}}}
+ \widetilde{F}^{\nu\eta\mu}_{{\bf k},{\bf q}{\bf p}} 
(u_{\nu{\bf k}}+v_{\nu{\bf k}})
(u_{\eta{\bf q}}v_{\mu{\bf p}}+v_{\eta{\bf q}}u_{\mu{\bf p}})\nonumber\\
&&\phantom{\widetilde{\Xi}^{\eta\nu\mu}_{{\bf q}{\bf k}{\bf p}}}
+ \widetilde{F}^{\mu\eta\nu}_{{\bf p},{\bf q}{\bf k}} 
(u_{\mu{\bf p}}+v_{\mu{\bf p}})
(u_{\eta{\bf q}}v_{\nu{\bf k}}+v_{\eta{\bf q}}u_{\nu{\bf k}})\, ,\nonumber\\
&&\widetilde{\Phi}^{\eta\nu\mu}_{{\bf q}{\bf k};{\bf p}} =  
\widetilde{F}^{\eta\nu\mu}_{{\bf q},{\bf k}{\bf p}} 
(u_{\eta{\bf q}}+v_{\eta{\bf q}})
(u_{\nu{\bf k}}u_{\mu{\bf p}}+v_{\nu{\bf k}}v_{\mu{\bf p}})
\label{eq_decayvertex}
\\
&&\phantom{\widetilde{\Phi}^{\eta\nu\mu}_{{\bf q}{\bf k};{\bf p}}}
+ \widetilde{F}^{\nu\mu\eta}_{{\bf k},{\bf p}{\bf q}} 
(u_{\nu{\bf k}}+v_{\nu{\bf k}})
(u_{\eta{\bf q}}u_{\mu{\bf p}}+v_{\eta{\bf q}}v_{\mu{\bf p}})\nonumber\\
&&\phantom{\widetilde{\Phi}^{\eta\nu\mu}_{{\bf q}{\bf k};{\bf p}}}
+ \widetilde{F}^{\mu\eta\nu}_{{\bf p},{\bf q}{\bf k}} 
(u_{\mu{\bf p}}+v_{\mu{\bf p}})
(u_{\eta{\bf q}}v_{\nu{\bf k}}+v_{\eta{\bf q}}u_{\nu{\bf k}}). \nonumber\quad\quad
\end{eqnarray}
We remark here that the final form of the three-magnon terms in (\ref{Hsource}) and (\ref{Hdecay})
and the formal expressions of the corresponding vertices in (\ref{eq_sourcevertex}) and (\ref{eq_decayvertex})
are virtually identical to the same expression for the kagome-lattice case, \cite{kagome15} with 
all  specifics of the problems hidden in the actual expressions for  $\widetilde{F}^{\mu\eta\nu}$ vertices and for the 
Bogolyubov parameters. In the present case,  summation over the  indices $\mu,\nu,\eta$ also
involves only two species of bosons instead of three in the kagome-lattice case. 

One of the key differences of the present case from the previously studied systems is in the complex
phase factors in (\ref{gk}), which originate 
from the lack of inversion symmetry with respect to the lattice points of the honeycomb lattice.
We point out that the symmetric distribution of these phase factors 
among boson operators in  (\ref{eq_phaseshift}) is responsible for the compact and purely real or purely
imaginary vertices $\widetilde{F}^{\mu\eta\nu}$ in (\ref{Fcompact}), which also translate 
to purely real or imaginary ultimate vertices in  (\ref{Hsource})-(\ref{eq_decayvertex}) and provide a 
significant advantage for both numerical evaluations of the decay rates and for analytical studies of their 
asymptotic behavior.

% ==============================================================================
\section{Kinematics of magnon decays}
% ==============================================================================
\label{sec_kinematics}
% ==============================================================================

In this Section, we discuss kinematic conditions for magnon decays, i.e. energy and momentum conservation
that define whether the nonlinear, anharmonic terms in the spin-wave Hamiltonian will lead to the broadening
in the magnon spectra.
The similarities and differences of this analysis, as compared to the previously studied models, are highlighted.

The phenomenon of spontaneous zero-temperature decays of quasiparticles has 
been studied for phonons in crystals, excitations in superfluid $^4\text{He}$, and for 
various types of bosonic excitations in quantum magnets.
\cite{RMP,pitaevskii,Zaliznyak06,Kosevich,Zheludev,YIG} 
The key consideration, common to all of the studied cases, is the 
determination of the conditions and ranges of the momenta that allow an elementary
two-particle decay processes to occur. This analysis can be performed independently of the actual calculation 
of the decay rates, based on the results of the harmonic approximation for the excitation energies, and
is simply expressed as  $\varepsilon_{\bf k}=\varepsilon_{\bf q}+\varepsilon_{\bf k-q}$ condition.
While one can expect spectrum renormalization effects beyond the harmonic theory 
and also contributions of the higher-order processes to play a role,
such a consideration is still immensely instructive. Not only does it help to identify the regions where decays
occur already in the lowest Born approximation and thus are likely to be the strongest even if 
the higher-order processes are included, but it also provides an information on the presence of 
singularities in the spectrum, associated with the Van Hove singularities in the two-particle continuum,
to which the single-particle branch is coupled.\cite{RMP}
 
A distinct feature of the present consideration is in having two branches of excitations, 
which makes multiple decay channels possible, so the general decay condition is modified to
\begin{eqnarray}
\varepsilon_{\mu{\bf k}}= \varepsilon_{\nu{\bf q}}+\varepsilon_{\eta{\bf k-q}},
\label{eq_channels}
\end{eqnarray}
where $\mu,\nu,\eta=1,2$  and we will denote a particular decay channel as $\mu\rightarrow\{\nu,\eta\}$. 
More specifically, since, as is mentioned above,
$\varepsilon_{2{\bf k}}\!>\!\varepsilon_{1{\bf k}}$ for any $\Delta<1$ at any value of the field, 
out of nominally six potential decay channels there are only three that can be kinematically 
allowed in the $XXZ$ honeycomb-lattice 
model in a field, one for the lower branch, $1\rightarrow\{1,1\}$, and two for the upper branch,
$2\rightarrow\{1,1\}$ and $2\rightarrow\{2,1\}$.

Another, more subtle distinction of the present case is in the existence of the Dirac-like degeneracy points between the 
two branches of excitations at the ${\rm K}$ and ${\rm K'}$  points of the Brillouin zone, which are also fixed at the energy 
$3JS$, independently of the value of the field or anisotropy.

% ==============================================================================
\subsection{Lower branch decays}
% ==============================================================================
% ==============================================================================
\begin{figure}[t]
\includegraphics[width=0.99\linewidth]{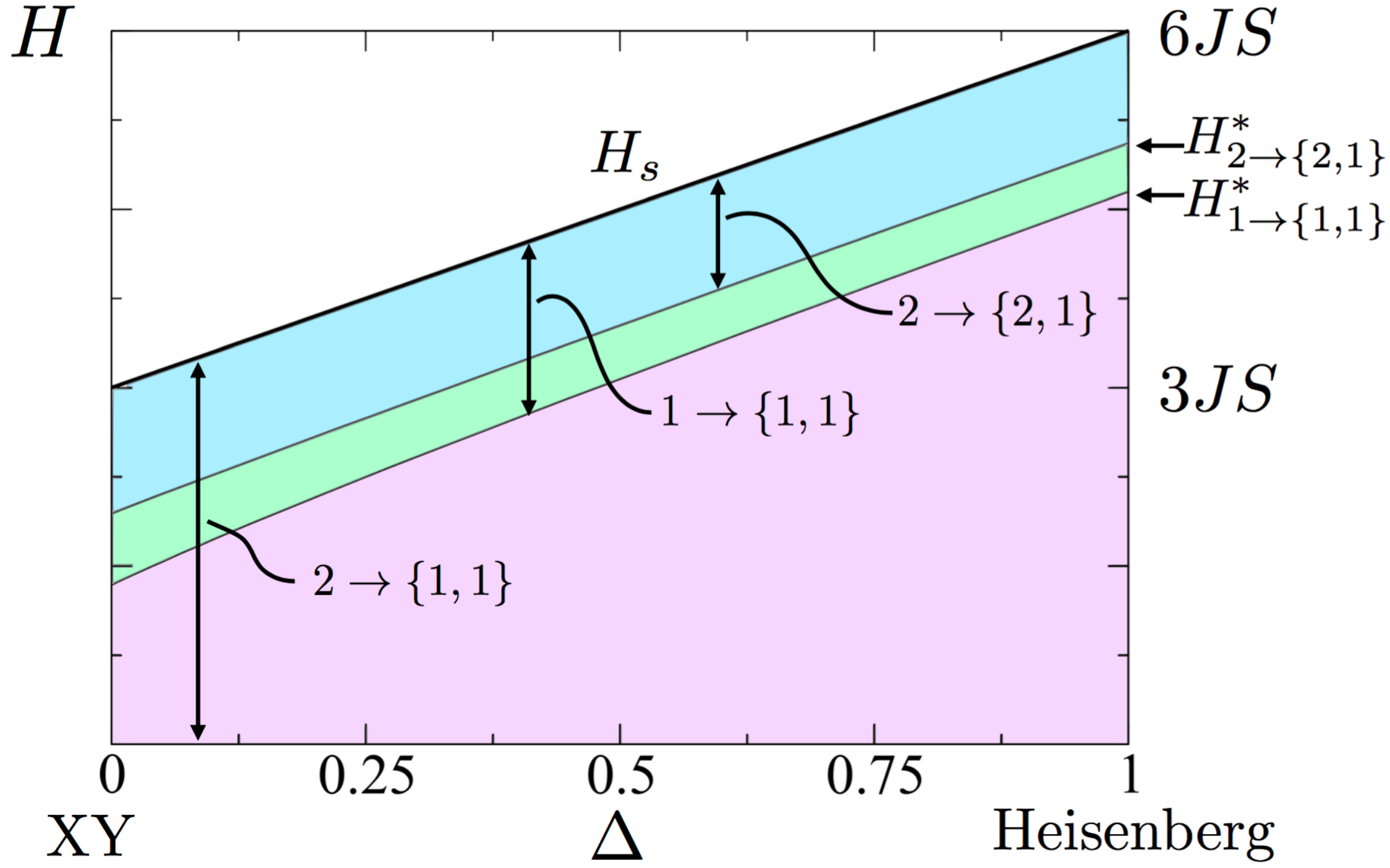}
\caption{(Color online)\ The $H$ vs $\Delta$ diagram, showing the range of fields for each $\Delta$
where different channels of  decays are allowed. Upper boundary is the saturation field
$H_s=3JS(1+\Delta)$. The spontaneous magnon decays in the lower branch are allowed between $H_s$ 
and $H^{*}_{1\rightarrow\{1,1\}}$ lines (\ref{H*111}), while 
decays in the $2\rightarrow \{2,1\}$ channel are allowed between $H_s$ and $H^{*}_{2\rightarrow\{2,1\}}$ line,
both indicated by vertical arrows. 
Decays in the $2\rightarrow\{1,1\}$ channel are possible for any $H_s>H>0$ and $1\geq\Delta\geq 0$; see text.}
\label{fig_phasediagram1}
\end{figure}
% ==============================================================================

The decay channel for the lower branch, $1\rightarrow\{1,1\}$, bears a lot of similarity to the square-lattice case,
although there are also some important differences. The main similarity is in the existence of a ``critical'' (threshold) 
value of the field  $H^{*}_{1\rightarrow\{1,1\}}$ at which decays become kinematically allowed.
This boundary is related to the change of the curvature of the long-wavelength part of the spectrum vs field, which is
also similar to the consideration given by Pitaevskii\cite{pitaevskii} to the phonon-like excitations in $^4$He. 
Considering the low-energy magnons and expanding $\varepsilon_{1{\bf k}}$ in (\ref{eq_energyspectrum}) 
near the $\Gamma$ point gives
\begin{equation}
\varepsilon_{1{\bf k}}\approx 3JS|{\bf k}|\sqrt\frac{1-\lambda}{6}
\left( 1+|{\bf k}|^2 \frac{5\lambda-3}{96(1-\lambda)} \right),
\label{w1_curvature}
\end{equation}
where $|{\bf k}|$ is in units of inverse lattice spacing $1/a$. 
A long-wavelength excitation becomes unstable towards spontaneous decays when the second term in the bracket 
is positive, corresponding to the positive, i.e. upward curvature of the spectrum. 
As is obvious from (\ref{w1_curvature}), in our case this happens for $\lambda>0.6$. 
Using the implicit relation between the field and anisotropy via (\ref{lambda}), 
a simple algebra yields
\begin{equation}
H^{*}_{1\rightarrow\{1,1\}}=H_s\sqrt{\frac{1+5\Delta}{5(1+\Delta)}}.
\label{H*111}
\end{equation}
Thus, the $1\!\rightarrow\!\{1,1\}$ decays are allowed in the range of fields 
$H^{*}_{1\rightarrow\{1,1\}}\!<\!H\!<\!H_s$,
shown in Fig.~\ref{fig_phasediagram1} vs $\Delta$. We note that at the saturation field, $H_s$, the decays 
{\it are} still kinematically allowed, but the fully saturated, ferromagnetic-like state forbids a direct
coupling of the single-magnon and the two-magnon sectors, thus forbidding the decays.\cite{RMP}

% ==============================================================================
\begin{figure}[t]
\frame{\includegraphics[width=0.9\linewidth]{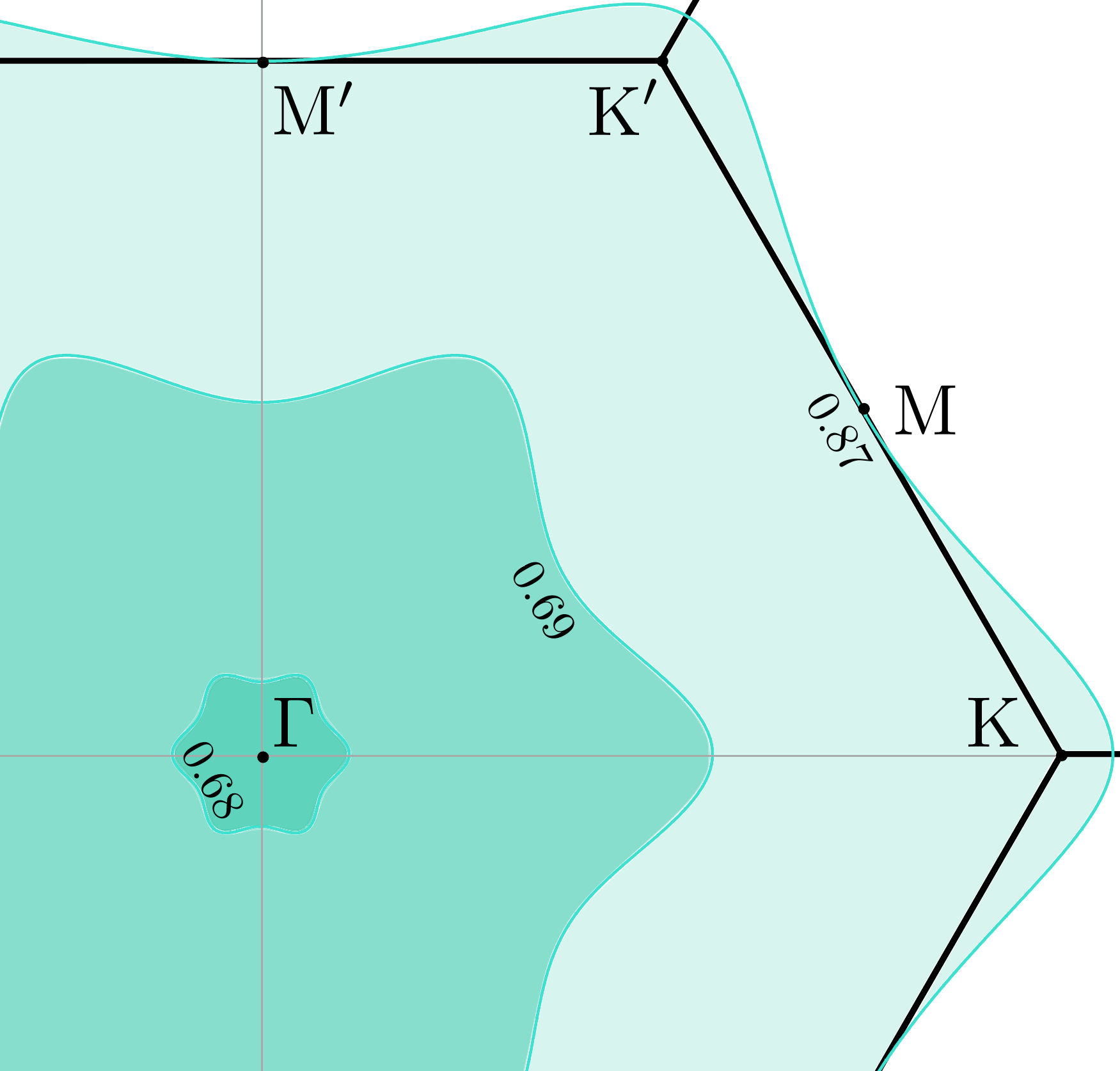}}
\caption{(Color online)\ The field-evolution of the decay regions (shaded) for the 
$1\!\rightarrow\!\{1,1\}$ decay channel in the first Brillouin zone for $\Delta=0.5$.
The numbers correspond to $H/H_s$. Decay threshold condition is $\lambda=0.6$ (see text). For different 
$\Delta$, the fields will change according to a relation in \eqref{lambda}.}
\label{fig_d1_region}
\end{figure}
\label{sec_lowerbranch}
% ==============================================================================

The next steps in the kinematic analysis  are the studies of the extent of the region in the Brillouin zone where
 decays are allowed for a given field and of the evolution of such a region vs field. 
These studies are  based on a simple observation that the region where decays in a given channel are allowed must be separated
from the region where such decays are forbidden by a \emph{decay threshold boundary},\cite{RMP} which, in turn, 
must correspond to a crossing of the single-particle excitation branch with the \emph{minimum} of the two-particle
continuum, e.g. $E_2 ({\bf k},{{\bf q}})=\varepsilon_{1{\bf q}}+\varepsilon_{1{\bf k-q}}$ for the considered 
$1\!\rightarrow\!\{1,1\}$ channel. 
Thus, the search for the decay threshold boundaries generally reduces to the search for the extrema of the 
two-particle continuum, determined from $\nabla_{{\bf q}} E_2 ({\bf k},{{\bf q}})=0$, which is satisfied 
when the two decay products have equal velocities, ${\bf v}_{{\bf q}}={\bf v}_{{\bf k-q}}$, where 
${\bf v}_{{\bf q}}=\nabla_{{\bf q}}\varepsilon_{{\bf q}}$. This equation must be solved together with the 
energy conservation condition (\ref{eq_channels}).
In fact, such a search yields more than just  decay threshold boundaries, as it also provides contours (in 2D) along 
which the single particle branch meets the other Van Hove singularities of the two-magnon
continuum, such as saddle points, which, generally, lead to  singularities in the decay rate, 
see Ref.~\onlinecite{RMP} and the next Section.

Some of the typical solutions for the decay threshold boundary are:\cite{RMP} 
(i)  emission of two magnons with equal momenta, ${\bf q}=\frac{1}{2}({\bf k+G}_i)$, 
where ${\bf G}_i$ is a reciprocal lattice vector, which gives an implicit solution for the contour in the form
 $\varepsilon_{1{\bf k}}=2\varepsilon_{1({\bf k+G}_i)/2}$, and 
(ii)  emission of an acoustic magnon, ${\bf q}\rightarrow 0$, which means 
that the velocity of the decaying magnon is equal to the velocity of the Goldstone mode
$|{\bf v}_{{\bf k}}|= v_0$.

In the considered  case of the $1\!\rightarrow\!\{1,1\}$ channel, the boundary of the decay region 
is determined by an emission of magnons with equal momenta, case (i) above,
for any value of the field. The field-evolution of the decay threshold boundary is shown in Fig.~\ref{fig_d1_region} 
for $\Delta=0.5$.  Decay region nucleates at the $\Gamma$ point at the 
field slightly exceeding $H \approx 0.68 H_s$ and grows quickly with the increasing field. 
At about $H \approx 0.87H_s$ there are no regions left in the Brillouin zone where magnons are stable.
Since our analysis is based on the the harmonic spectrum, which 
depends on $\Delta$ and $H$ via a single parameter $\lambda$ (\ref{lambda}),
a similar field-evolution of the decay region boundary is expected for all other values of $\Delta$.

We note the differences of these results from the case of the square-lattice antiferromagnet.\cite{RMP,99,Mourigal10}
In the square-lattice case, the decay boundary at $H\agt 0.8H_s$ is partially determined by the emission 
of the acoustic magnon, case (ii) above. Another important difference is that while in the present case the Brillouin zone is 
fully taken over by the decays at $H<H_s$, in the 
square-lattice case the decay region reaches all  corners of the Brillouin zone only asymptotically
at $H\rightarrow H_s$. This difference originates in the structure
of the high-energy portion of the lower-branch spectrum that has Dirac-like ${\rm K}$-points at
fixed energy, see Fig.~\ref{fig_spectrum}, which are not true extrema, but have a cone-like dispersion in their 
vicinity.

% ==============================================================================
\subsection{Upper branch decays}
% ==============================================================================
\label{sec_upbranchkin}
% ==============================================================================

In the considered case of the honeycomb lattice  we have two branches of excitations. 
While magnons from the lower, acoustic branch are only able to decay into themselves, the upper-branch optical 
magnons have two channels of decay that should be analyzed separately.

% ==============================================================================
\subsubsection{$2\rightarrow \{2,1\}$ decay channel}
% ==============================================================================

First of the two is the decay of the optical magnon into  the optical  and  acoustic ones, denoted  as the
$2\rightarrow \{2,1\}$ decay channel.
Similarly to the acoustic branch decays discussed above, the kinematic conditions for this channel 
are met only above a threshold field value, see Fig.~\ref{fig_phasediagram1}. 
By inspection, the first type of the decay processes encountered by the optical branch upon 
the field increase is the emission of the Goldstone mode of the acoustic branch, similar to the
case (ii) of the previous consideration. 

% ==============================================================================
\begin{figure}[t]
\frame{\includegraphics[width=0.9\linewidth]{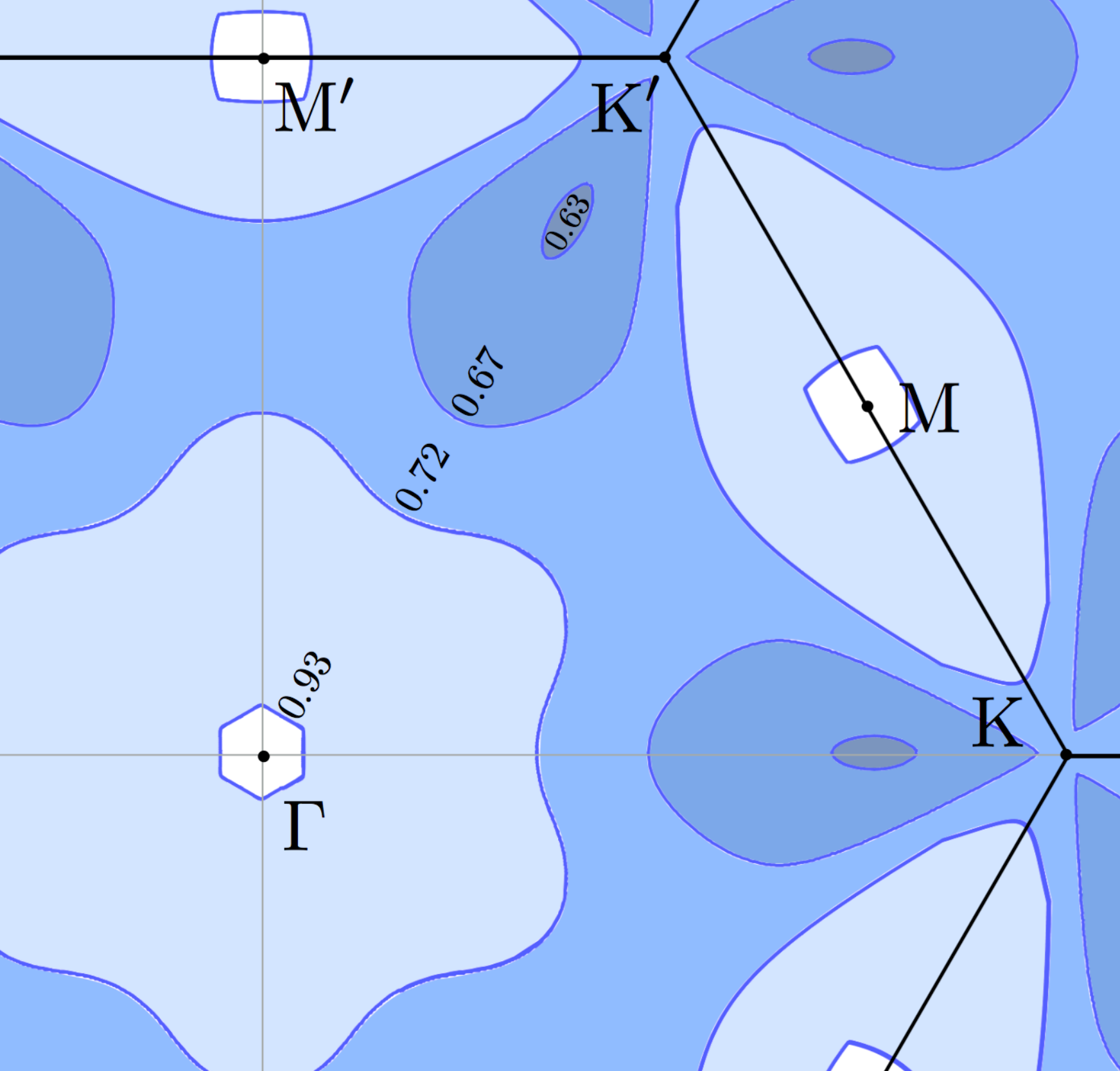}}
\caption{(Color online) 
\ The field-evolution of the decay regions (shaded) for the 
$2\rightarrow \{2,1\}$ decay channel   for $\Delta=0$, numbers are  $H/H_s$.
Decays nucleate at $H\simeq 0.63 H_s$ ($\lambda=0.707$) in the vicinity 
of the ${\rm K}$-points and grow with increasing field with the  
decay-free regions disappearing  at $H\rightarrow H_s$. A peculiar shape of the decay boundaries at larger fields hints at
several types of the decay thresholds controlling them; see text. For different 
$\Delta$, the fields change according to \eqref{lambda}.}
\label{fig_d122_regions2}
\end{figure}
% ==============================================================================

The first instance when such a decay is possible is when the velocity of the optical magnon anywhere in the Brillouin zone
exceeds the velocity of the $|{\bf q}|\rightarrow 0$ mode of the acoustic branch: 
$|{\bf v}_{2{\bf k}}|\!\geq v_{1}\!=\!3JS\sqrt{(1-\lambda)/6}$. 
Na\"{i}vely, this condition should be first met when the slope of the Dirac-like cone for the upper branch at the ${\rm K}$-point
matches the slope of the Goldstone mode, see Fig.~\ref{fig_spectrum} for  guidance.
However, the Dirac cone has an anomalous dispersion, in which the linear term is followed by the 
$|{\bf k}|^2$-term with the varying convexity, depending on the azimuthal angle as $\sim\cos 3\varphi$.
Because of that,  matching of the velocities first occurs at a finite distance from the  ${\rm K}$-points and at 
particular angles. Our Figure~\ref{fig_d122_regions2} demonstrates this feature for 
a representative value of $\Delta=0$, for which the threshold field is about $H^{*}_{2\rightarrow\{2,1\}}=0.63 H_s$.

A somewhat tedious but straightforward expansion of the energies in \eqref{eq_energyspectrum}
for small $|{\bf k}|$ near the ${\rm K}$-point in the ${\rm K}\rightarrow\Gamma$ direction, 
keeping  terms up to $|{\bf k}|^3$, followed by the maximization of the velocity $|{\bf v}_{2{\bf k}}|$
yields an implicit equation for the threshold value of $\lambda^*$
\begin{equation}
\frac{\lambda}{2}+\frac{1}{12}
\left(\frac{(3\lambda-1-\lambda^2)^2}{1-2\lambda+3\lambda^2}\right)
=\sqrt{\frac{1-\lambda}{2}}\, ,
\label{lambda*}
\end{equation}
with an approximate solution $\lambda^*\!\approx \!0.707$. To obtain the threshold field vs $\Delta$
one needs to resolve  (\ref{lambda}), which yields
$H^{*}_{2\rightarrow\{2,1\}}/H_s\!=\!\sqrt{(\Delta+2\lambda^*-1)/(1+\Delta)}$. 
Verifying this approximate answer (\ref{lambda*}) for $\Delta\!=\!0$ 
gives $H^{*}_{2\rightarrow\{2,1\}}\!\approx\! 0.64 H_s$, close to the numerical value in 
Figs.~\ref{fig_d122_regions2} and \ref{fig_phasediagram1}.
A location of the velocity maximum  that corresponds to the nucleation point of  decays in this channel
can also be found from the same approach to be at
\begin{equation}
|{\bf k}^*|=\frac{2}{\sqrt{3}}\left(\frac{3\lambda-1-\lambda^2}{1-2\lambda+3\lambda^2}\right)
\Big|_{\lambda^*} %\approx 0.66
\approx 0.158|{\bf K}|,
\end{equation}
which also compares favorably with the numerical result $|{\bf k}^*|\!\approx \!0.23|{\bf K}|$ 
in Fig.~\ref{fig_d122_regions2}.

\begin{figure}[t]
\frame{\includegraphics[width=0.9\linewidth]{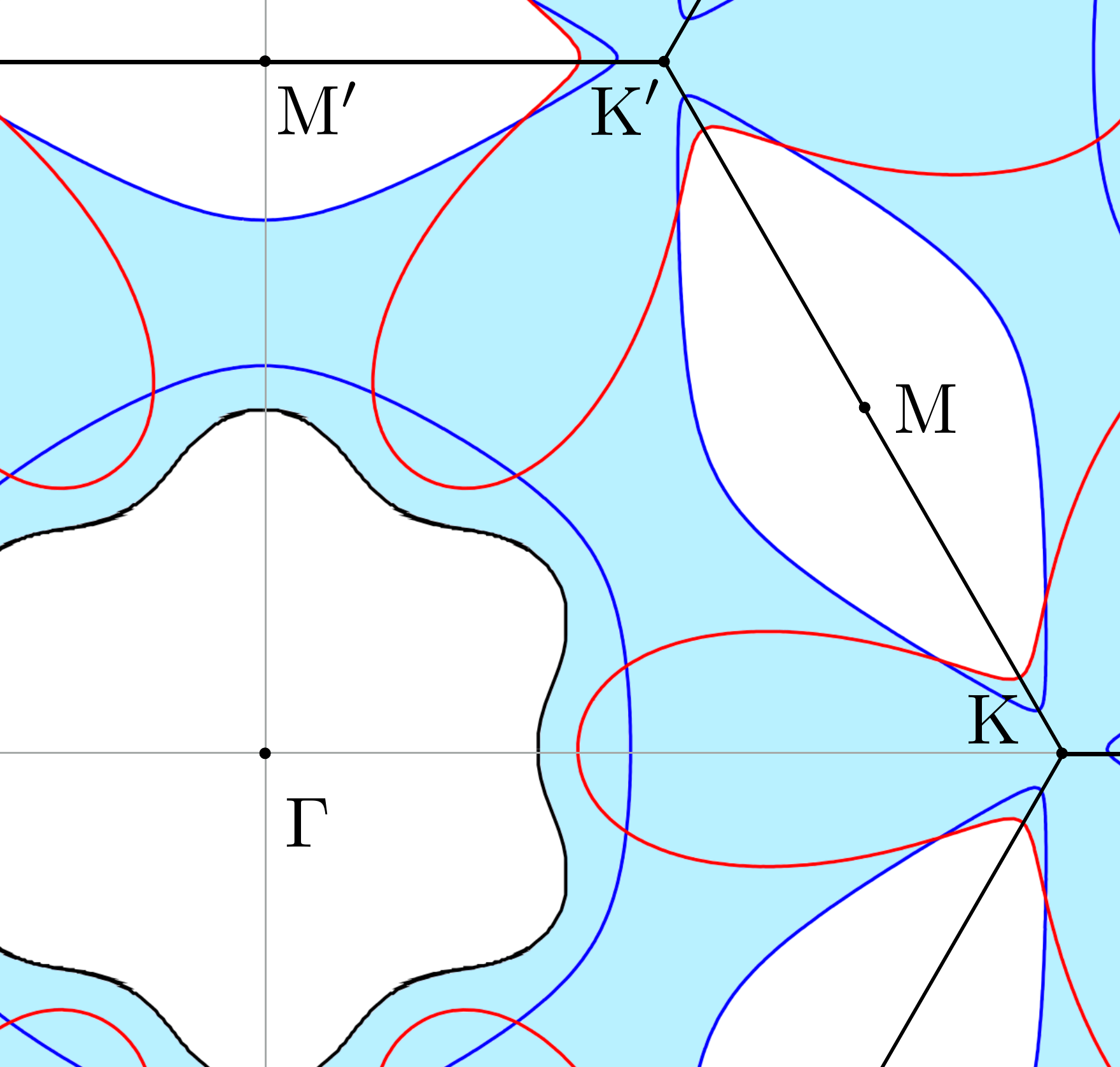}}
\caption{(Color online)\ The decay region for the $2\rightarrow \{2,1\}$ channel (shaded)  for $\Delta\!=\!0$ and 
$H/H_s\!=\!0.72$. [Same holds for any $H$ and $\Delta$ satisfying $\lambda=0.76$ via \eqref{lambda}]. 
Lines are contours for the different types of thresholds.
In the $\Gamma$ to ${\rm M}$ direction, the second and the third contours 
correspond to the emission of the Goldstone magnon from acoustic branch. 
The closest contour  around the $\Gamma$ point is a generic threshold boundary for the
emission of two finite-energy magnons with different energies but same velocities.
Contours  enclosing the ${\rm K}$-points are thresholds for the Dirac-mode emission; see text.}
\label{fig_d122_regions1}
\end{figure}

There are two notable features of the field-evolution of the decay regions shown in  Fig.~\ref{fig_d122_regions2}.
First, unlike in the $1\rightarrow\{1,1\}$ channel, the decay-free regions of the Brilloiun zone 
are  eliminated completely only at the saturation field $H\rightarrow H_s$. 
Second, the evolution of the decay threshold boundaries is far more intriguing.
Upon increase of the field, only part of the decay boundaries is defined by  
the Goldstone emission, as can be seen in the more peculiar shapes of the boundaries.
This is analyzed in more detail in Fig.~\ref{fig_d122_regions1} 
for $\Delta=0$ and $H=0.72H_s$, which shows a union of three types of boundaries. 

% ==============================================================================
% ==============================================================================

First type contains two contours
that are identified with the original case of the Goldstone mode emission. These are the second and the third contours 
to cross if traversing from the $\Gamma$ to ${\rm M}$ point.
The second type is the first contour that surrounds the $\Gamma$-point. It is a generic threshold boundary for the
emission of two finite-energy magnons with different energies but same velocities, which cannot be simplified 
to   typical cases   (i) or (ii) considered above. The last type consists of a set of contours 
having elongated shapes that are enclosing the ${\rm K}$-points.
It corresponds to a different type of a solution for the two-magnon continuum energy minimum, 
which has not been discussed previously. Instead of a Goldstone-mode emission, one can have
a Dirac-mode emission in the presence of  Dirac-like cones in the spectrum, i.e. 
when one of the decay products is a particle at one of the ${\rm K}$-points
\begin{equation}
\varepsilon_{2{\bf k}}=\varepsilon_{2{\bf K}}+\varepsilon_{1{\bf k-K}}\, ,
\label{min_dirac}
\end{equation}
where ${\bf v}_{2{\bf K}}\!=\!{\bf v}_{1{\bf k-K}}$, as before.
Aside from serving as a decay threshold boundary for a part of the ${\bf k}$-space near the ${\rm K}$-points
around  the ${\rm M}\!\rightarrow\!{\rm K}$ direction in Fig.~\ref{fig_d122_regions1}, these Dirac-emission
threshold contours correspond to strong singularities in the decay rate 
discussed in the next Section. This is contrary to the 
Goldstone-emission contours, which are  associated with the vanishing density of states and  
correspond only to weak singularities. \cite{triangle,RMP} 

% ==============================================================================
\subsubsection{$2\rightarrow \{1,1\}$ decay channel}
% ==============================================================================

% ==============================================================================
\begin{figure}
\frame{\includegraphics[width=0.9\linewidth]{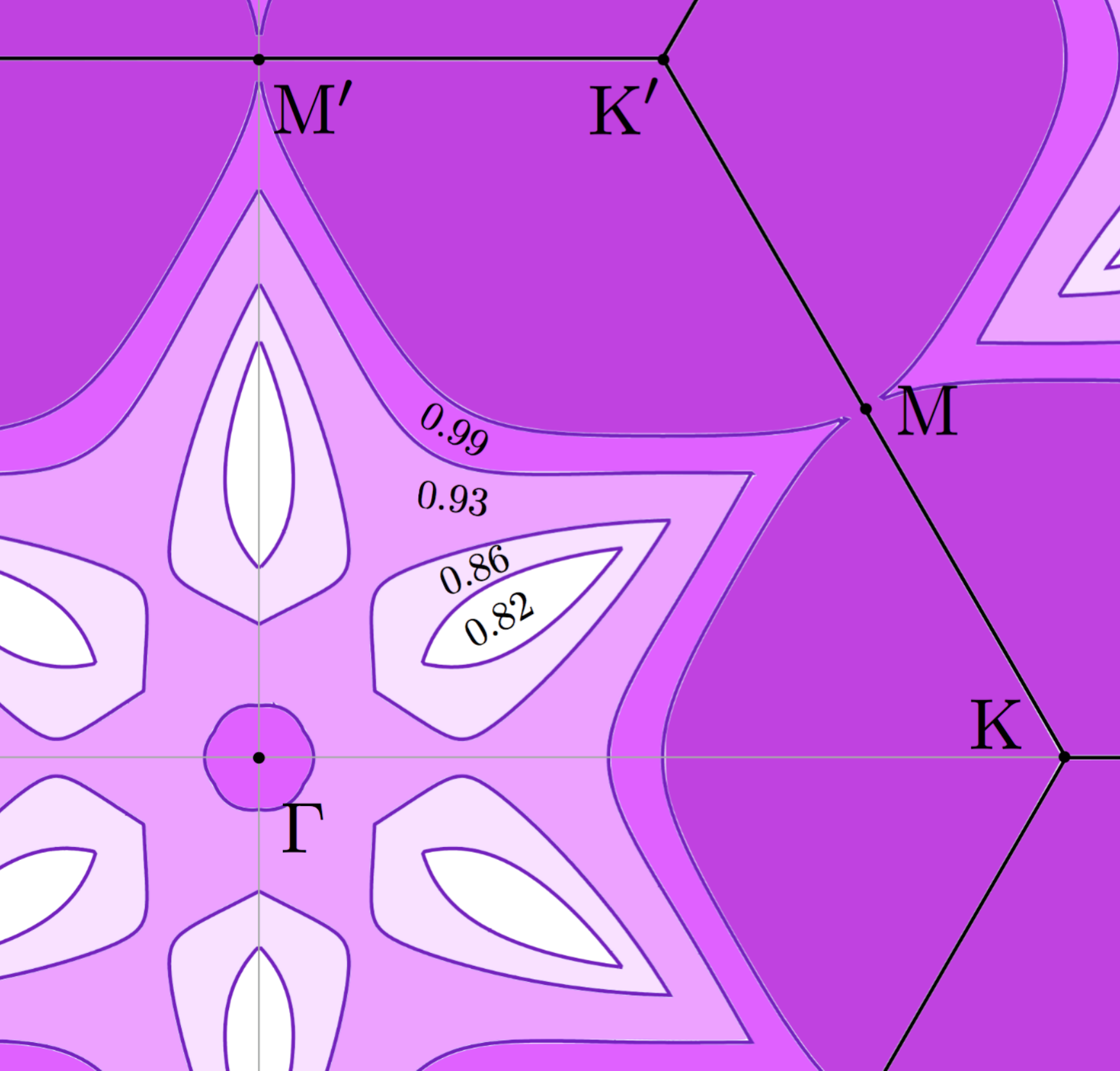}}
\caption{(Color online)\ 
The field-evolution of the decay regions of $2\!\rightarrow\! \{1,1\}$ channel (shaded) for $\Delta\!=\!0$, 
numbers are  $H/H_s$.
The decay-free regions nucleate at $H\simeq 0.82 H_s$ ($\lambda=0.81$) 
in the vicinity of  $\Gamma$ point and grow with 
increasing field. However, even at $H\!=\!H_s$ there are finite decay regions around ${\rm K}$ points. 
For different 
$\Delta$, the fields change according to \eqref{lambda}.}
\label{fig_d112_regions_evolution}
\end{figure}
% ==============================================================================

The last decay channel is distinct from the previously studied ones in that 
there are always regions of the Brillouin zone where decays are kinematically allowed for any $\Delta$ and $H$
except for $\Delta=1$ and $H=0$ point where magnon bands are degenerate. Otherwise, 
decays from the upper branch to the lower branch are always possible. 
This is true for \emph{all} the momenta in the Brillouin zone for $\lambda\alt 0.81$. 
However, at $\lambda\agt 0.81$, which corresponds to large field values, there are decay-free regions of the 
Brillouin zone. That is, the situation is somewhat inverse to the previously discussed channels, because instead 
of the nucleation of the decay regions upon increase of the field we have the nucleation of the decay-free regions.
With increasing field  the decay-free regions grow, although even for $H\rightarrow H_s$ they  occupy only part of the 
Brillouin zone, see Fig.~\ref{fig_d112_regions_evolution} where the evolution of the decay-free regions vs $H$ is shown 
for $\Delta=0$. 

% ==============================================================================
\begin{figure}
\frame{\includegraphics[width=0.9\linewidth]{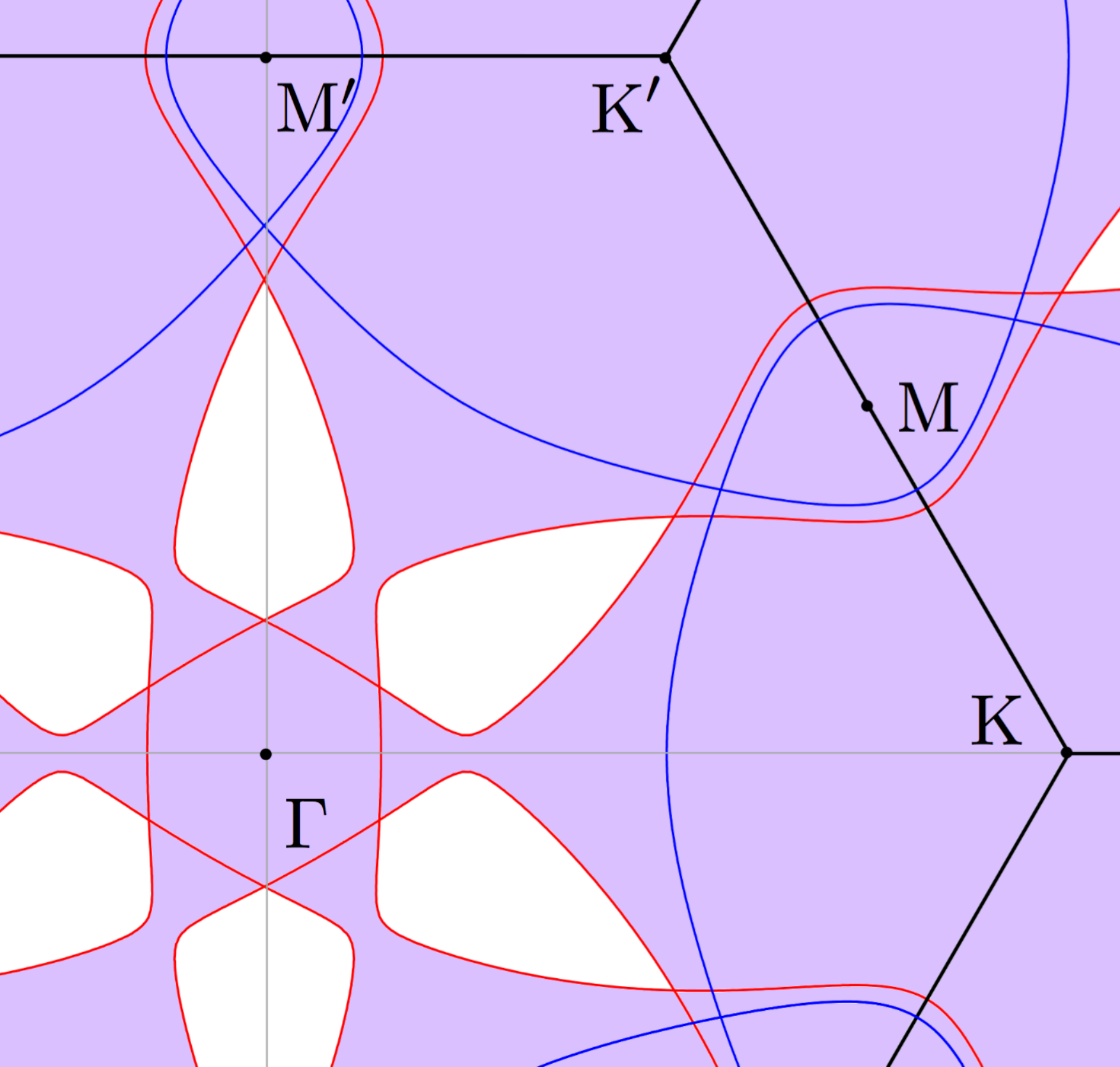}}
\caption{(Color online)\ The decay region of the $2\rightarrow \{1,1\}$ channel (shaded) 
and associated threshold contours for $\Delta=0$ and $H/H_s=0.86$. 
[Same holds for any $H$ and $\Delta$ satisfying $\lambda=0.86$ via \eqref{lambda}].  
Contours enclosing ${\rm K}$ points
correspond to decays into magnons with equal momenta, case (i). 
Contours, which contain the boundaries for the decay-free regions (not shaded), 
correspond to the Dirac-mode emission (\ref{min_dirac1}).}
\label{fig_d112_regions}
\end{figure}
% ==============================================================================

Still, the corresponding decay boundaries require the same threshold analysis.
One can show that the decay boundary is defined by the minimum of the two-magnon continuum 
that corresponds to an emission of a Dirac-mode from the acoustic branch at the ${\bf K}$-point
\begin{equation}
\varepsilon_{2{\bf k}}=\varepsilon_{1{\bf K}}+\varepsilon_{1{\bf k-K}},
\label{min_dirac1}
\end{equation}
which is, actually, the same process as in \eqref{min_dirac} because of the Dirac-point 
degeneracy of the magnon branches,
$\varepsilon_{1{\bf K}}=\varepsilon_{2{\bf K}}=3JS$.

In Fig.~\ref{fig_d112_regions}, we show two different  threshold contours for the 
$2\rightarrow \{1,1\}$ decay channel at a representative $\Delta=0$ and $H=0.86H_s$.
Both contours should be associated with the Van Hove singularities in the two-magnon continuum.
The first set of contours, parts of which serve as the boundaries for the decay-free regions (not shaded), 
correspond to the Dirac-mode emission (\ref{min_dirac1}). The other contours, enclosing the ${\rm K}$-points,
correspond to  decays with an emission of magnons with equal momenta, case (i) discussed above.

% ==============================================================================
\section{Magnon decays}
% ==============================================================================
\label{sec_cubic}
% ==============================================================================

The most important \emph{qualitative} differences of the spectral properties of the noncollinear magnets
vs their collinear counterparts occur due to anharmonic cubic coupling terms in the spin Hamiltonians 
of the type (\ref{Hdecay}) that may induce magnon decays.\cite{RMP} 
Concomitant to the decays are the singularities in the spectrum that \emph{necessarily} occur 
due to Van Hove singularities within the two-magnon continua via the coupling of the latter to the 
single-particle branch mediated by the same cubic terms. 

In addition to the purely kinematic analysis provided in Sec.~\ref{sec_kinematics} above, which has 
identified decay thresholds and regions for all the decay channels relevant to our model, we now analyze
the actual magnon decay rate. The goal is to identify the field regimes when  decays are particularly
strong, find asymptotic behavior of the damping in the proximity of the high-symmetry points, 
and search for unusual features due to singularities, altogether providing a  guidance for experimental
fingerprinting of decay-induced spectral features.
 
Using standard diagrammatic rules with the decay vertices in \eqref{Hdecay} we obtain  the
decay rate  in the lowest Born approximation in the $\mu\! \rightarrow\! \{ \eta, \nu \}$ channel
as given by
\begin{eqnarray}
\Gamma^{\mu \rightarrow \{ \eta, \nu \}}_{{\bf k}} =
\frac{\pi}{2} \sum_{{\bf q}} \left| \Phi^{\eta\nu\mu}_{{\bf q},{\bf k-q};{\bf k}}\right|^2 
\delta(\varepsilon_{\mu{\bf k}}-\varepsilon_{\eta{\bf q}}-\varepsilon_{\nu{\bf k-q}}).
\label{eq_gammak}
\end{eqnarray}
As  discussed  in Sec.~\ref{sec_kinematics}, there are three relevant decay channels for the  
$XXZ$ model on the honeycomb lattice in a field. 
For the two of them, $\mu \rightarrow \{ 1, 1 \}$ with $\mu=1$ or $2$, the decay rate can be written replacing
vertices and energies by their dimensionless counterparts as
\begin{eqnarray}
&&\Gamma^{\mu \rightarrow \{ 1, 1 \}}_{{\bf k}} = \Gamma_{\theta,\Delta}  \sum_{{\bf q}} 
\left| \widetilde{\Phi}^{11\mu}_{{\bf q},{\bf k-q};{\bf k}}\right|^2\nonumber\\
&&\phantom{\Gamma^{\mu \rightarrow \{ 1, 1 \}}_{{\bf k}} = \Gamma_{\theta,\Delta}  \sum_{{\bf q}}}
\times\delta \left(\omega_{\mu{\bf k}}-\omega_{1{\bf q}}-\omega_{1{\bf k-q}}\right),\ \ \
\label{eq_gamma11}
\end{eqnarray}
where we have introduced an auxiliary  constant
\begin{eqnarray}
\Gamma_{\theta,\Delta} = \frac{\pi}{6JS} \left| J^{(3)}_{\theta,\Delta}\right|^2  =\frac{3\pi}{16}\, J\, \left(1+\Delta\right)^2
\sin^22\theta\, .
\label{eq_gamma0}
\end{eqnarray}
The decay rate in the  $2 \rightarrow \{ 2, 1 \}$ channel has an additional factor of two because of the permutation 
of the acoustic and optical decay products
\begin{eqnarray}
&&\Gamma^{2 \rightarrow \{ 2, 1 \}}_{{\bf k}} = 2\Gamma_{\theta,\Delta}  \sum_{{\bf q}} 
\left| \widetilde{\Phi}^{122}_{{\bf q},{\bf k-q};{\bf k}}\right|^2\nonumber\\
&&\phantom{\Gamma^{2 \rightarrow \{ 2, 1 \}}_{{\bf k}} = \Gamma_{\theta,\Delta}  \sum_{{\bf q}}}
\times\delta \left(\omega_{2{\bf k}}-\omega_{2{\bf q}}-\omega_{1{\bf k-q}}\right).\ \ \
\label{eq_gamma122}
\end{eqnarray}
It is worth to note that since  spin canting is induced by  external magnetic field, the field 
is crucial to the phenomenon of magnon decays in the studied system. 
This is clear from the explicit form of $\Gamma_{\theta,\Delta}$ in (\ref{eq_gamma0}), which vanishes  at $\theta=0$, see also
Fig.~\ref{fig_lattice}. Another vanishing point is  $\theta=\pi/2$, which corresponds to the saturation
field. Thus, in both limits  decay vertex is zero and
while magnon decays may be kinematically allowed per our discussion in Sec.~\ref{sec_kinematics} above,
the absence of a coupling between single- and two-magnon sectors renders decays impossible.

% ==============================================================================
\begin{figure*}
\includegraphics[width=1\textwidth]{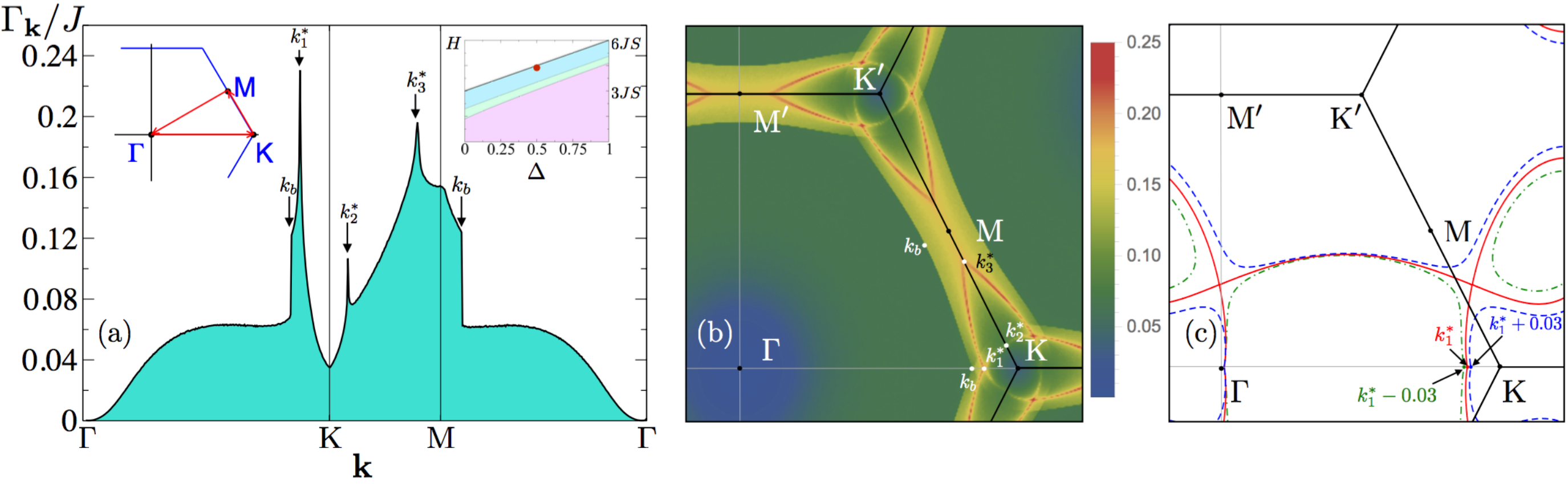} 
\caption{(Color online)\
$\Gamma^{1\rightarrow \{ 1, 1 \}}_{{\bf k}}$ vs ${\bf k}$ calculated using (\ref{eq_gamma11}) for 
$\Delta\!=\!0.5$ and $H\!=\!0.98H_s$, (a) along the 
$\Gamma\!\rightarrow\!{\rm K}\!\rightarrow\!{\rm M}\!\rightarrow\!\Gamma$ path, 
and (b) intensity plot across the Brillouin zone. Right inset in (a)  shows the chosen $\{\Delta,H\}$ point in the
diagram Fig.~\ref{fig_phasediagram1}. Arrows indicate singularities discussed in text. 
(c) Decay contours in the ${\bf q}$-space 
for a magnon at  $k_1^{*}-0.03$ (dashed-dotted), $k_1^{*}$ (solid), and $k_1^{*}+0.03$ (dashed)  
[see (a,b) for $k_1^{*}$]. The decay contours undergo a topological transition upon crossing 
$k_1^*$; see text.}
\label{fig_gamma1}
\end{figure*}
% ==============================================================================

% ==============================================================================
\subsection{Decays of the lower branch magnons}
% ==============================================================================

As  mentioned in Sec.~\ref{sec_kinematics},  decays of magnons in the acoustic branch
are only into the products that belong to the same species and the field evolution of these decays 
bears a significant similarity to the case of the square-lattice antiferromagnet considered before.\cite{RMP}
Our Fig.~\ref{fig_gamma1}(a) shows $\Gamma^{1\rightarrow \{ 1, 1 \}}_{{\bf k}}$ vs ${\bf k}$ 
along the $\Gamma\!\rightarrow\!{\rm K}\!\rightarrow\!{\rm M}\!\rightarrow\!\Gamma$ cut of the Brillouin zone
for a representative $H\!=\!0.98H_s$ and $\Delta\!=\!0.5$. 
Here and in the following, the numerical Monte Carlo integration method was used in
\eqref{eq_gamma11} and \eqref{eq_gamma122} with $10^8$-$10^9$ integration points in the full Brillouin zone with
an artificial broadening of the $\delta$-function $\epsilon\!=\!5\!\times\! 10^{-4}$.
As discussed above, the field must exceed
$H^{*}_{1\rightarrow\{1,1\}}$ for  decays to become kinematically allowed.
One similarity with the square-lattice case is the asymptotic behavior of $\Gamma_{{\bf k}}$ in the
long-wavelength limit: $\Gamma^{1\rightarrow \{ 1, 1 \}}_{{\bf k}}\!\propto\!|{\bf k}|^3$ 
near the $\Gamma$ point in agreement with a general hydrodynamic expectations,\cite{RMP} 
see also  discussion of Fig.~\ref{fig_asymp1}(a). 

% ==============================================================================
\begin{figure}
\includegraphics[width=0.9\linewidth]{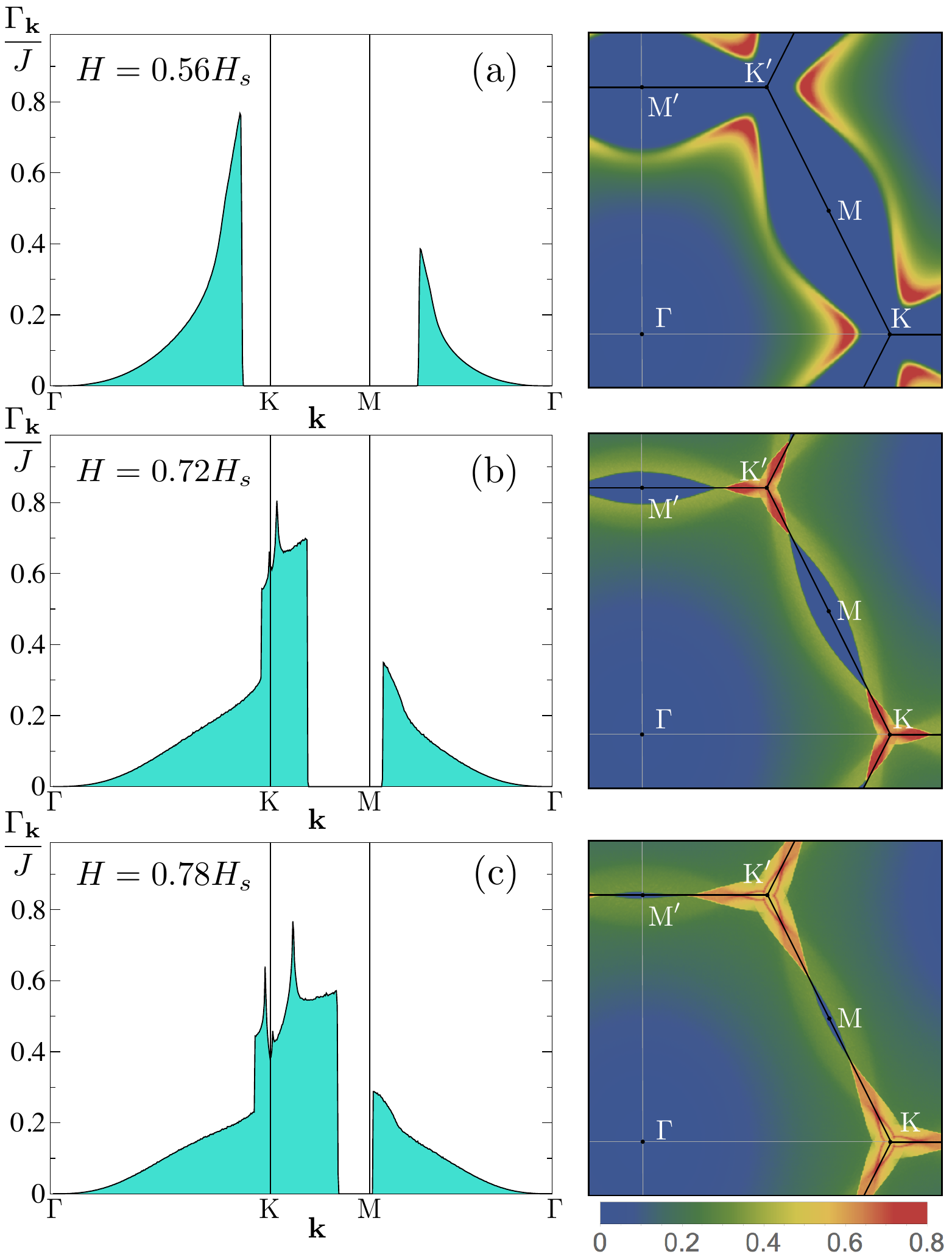}
\caption{(Color online)\ The field-evolution of $\Gamma^{1\rightarrow \{ 1, 1 \}}_{{\bf k}}$ vs ${\bf k}$ 
along the  $\Gamma\!\rightarrow\!{\rm K}\!\rightarrow\!{\rm M}\!\rightarrow\!\Gamma$ path (left), 
and intensity plot (right) for $\Delta=0$. (a) $H\!=\!0.56H_s$, (b) $H\!=\!0.72H_s$, and (c) $H\!=\!0.78H_s$, 
see also Fig.~\ref{fig_d1_region}. Threshold-boundary singularities persist for all fields, while saddle-point singularities
appear upon the overlap of the decay regions; see text.}
\label{fig_decay1e2}
\vskip -0.4cm
\end{figure}
% ==============================================================================

% ==============================================================================
\begin{figure*}
\includegraphics[width=\linewidth]{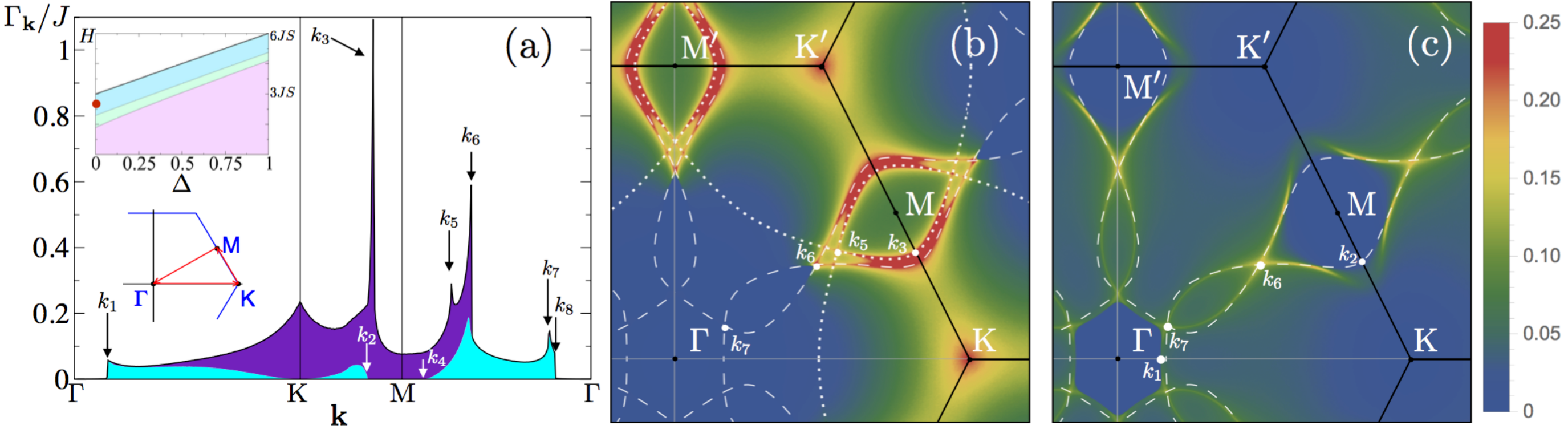}
\caption{(Color online)\
The decay rate of the optical branch for $\Delta\!=\!0$ and $H\!=\!0.84H_s$ 
[see upper inset in (a)] obtained from \eqref{eq_gamma11} and \eqref{eq_gamma122}, (a)
along the $\Gamma\!\rightarrow\!{\rm K}\!\rightarrow\!{\rm M}\!\rightarrow\!\Gamma$ path and
the 2D intensity maps of the (b) $\Gamma^{2 \rightarrow \{ 1, 1 \}}_{{\bf k}}$ and (c) 
$\Gamma^{2 \rightarrow \{ 2, 1 \}}_{{\bf k}}$ channels.
In (a), the upper and the lower shaded areas are    contributions 
of the $2\rightarrow \{1,1\}$ and $2\rightarrow \{2,1\}$ channels, respectively.
Arrows indicate representative singularities, contours in (b) and (c) are the thresholds 
of the decay into magnons with the same momenta (dotted) 
and to the emission of a Dirac-like magnon (dashed); 
see Sec.~\ref{sec_upbranchkin}, Figs.~\ref{fig_d122_regions1} and
\ref{fig_d112_regions}, and discussion in text. }
\label{fig_gamma2d0t1}
\end{figure*}
% ==============================================================================

The magnon decay rate in Figs.~\ref{fig_gamma1}(a,b) exhibits a number of singularities, which can be traced
back to the discussion of the threshold contours in Sec.~\ref{sec_kinematics}. First type of  singularities is the 
step-function-like behavior at the points denoted as $k_b$. They correspond to the decay threshold
boundaries due to decays into pairs of  magnons with equal momenta, 
which are associated with the minima in the two-magnon continuum, hence the step-like Van Hove 
singularities in 2D. A more detailed analysis of the field-evolution of $\Gamma^{1\rightarrow \{ 1, 1 \}}_{{\bf k}}$
is offered in Fig.~\ref{fig_decay1e2}, which shows that, initially, i.e. at smaller fields, these types of
singularities are the only ones present in the structure of $\Gamma^{1\rightarrow \{ 1, 1 \}}_{{\bf k}}$, but,
upon the field increase, the contours from the neighboring  Brillouin zones overlap 
and precipitate new types of singularities, see also Fig.~\ref{fig_d1_region}. 
However, the threshold boundaries from different Brillouin zones retain their step-like
character, see Figs.~\ref{fig_gamma1}(a) and \ref{fig_decay1e2}(b,c).

The second type of singularities in Figs.~\ref{fig_gamma1}(a,b) and \ref{fig_decay1e2}(b,c) that are characteristic
to  magnon decay phenomena, are the logarithmic singularities denoted as $k^{*}_i$ in Fig.~\ref{fig_gamma1},
which are associated with the saddle points in the two-magnon continuum.\cite{triangle,RMP}
These singularities appear in the single-magnon spectrum along the contours in ${\bf k}$-space that are the 
intersects of the surfaces of these saddle points with $\varepsilon_{\bf k}$ surface of the magnon branch.

An interesting characteristics of these singularities is that they correspond to topological transitions 
in a different set of contours, the so-called \emph{decay contours}, a set of ${\bf q}$-points at which the decay products are
created. An example of such a transition is shown in Fig.~\ref{fig_gamma1}(c) for the same set of parameters
as  in Fig.~\ref{fig_gamma1}(a,b).
It shows three sets of contours, upon the approach to, at, and upon the crossing of the $k^*_1$ point,
see Fig.~\ref{fig_gamma1}(a). 
As was discussed previously, such topological transitions are not only common, but in some cases can be shown to
necessarily exist based on the analysis of the structure of the ${\bf q}$-space manifold for the decay 
products.\cite{triangle,RMP}

The following remark concerns all channels of  decay analyzed in this work. 
Our consideration of $\Gamma_{\bf k}$ is limited to the $1/S$ effects, 
which corresponds to the one-loop Born approximation. The divergent singularities are regularized 
by the higher-order processes, such as cascade-type decays, yielding the finite lifetime in the
decay products that cuts off singularities. Another type of regularization in the case when decay products remain
stable are of a more complicated nature, see Refs.~\onlinecite{triangle,RMP} for a discussion. However,
we note that although the divergences in $\Gamma_{\bf k}$ are regularized, the characteristic magnitude 
and the overall shapes of $\Gamma_{\bf k}$ vs ${\bf k}$ can be expected to remain the same 
as was demonstrated previously by implementing different self-consistency schemes.\cite{Mourigal10,99,triangle,RMP}

% ==============================================================================
\subsection{Decays of the upper branch magnons}
% ==============================================================================

We proceed with the analysis of the decays in the upper branch of spin excitations.
As discussed in Sec.~\ref{sec_upbranchkin}, there are two potential decay channels for the optical branch,
into two lower branch magnons, $2\rightarrow \{1,1\}$, and into one optical magnon with an emission of 
acoustic one, $2\rightarrow \{2,1\}$. While the former is kinematically allowed for any $\Delta<1$ or $H>0$, for the
latter channel there is a threshold field  $H^{*}_{2\rightarrow\{2,1\}}$, see Fig.~\ref{fig_phasediagram1}.

\begin{figure}
\includegraphics[width=0.9\linewidth]{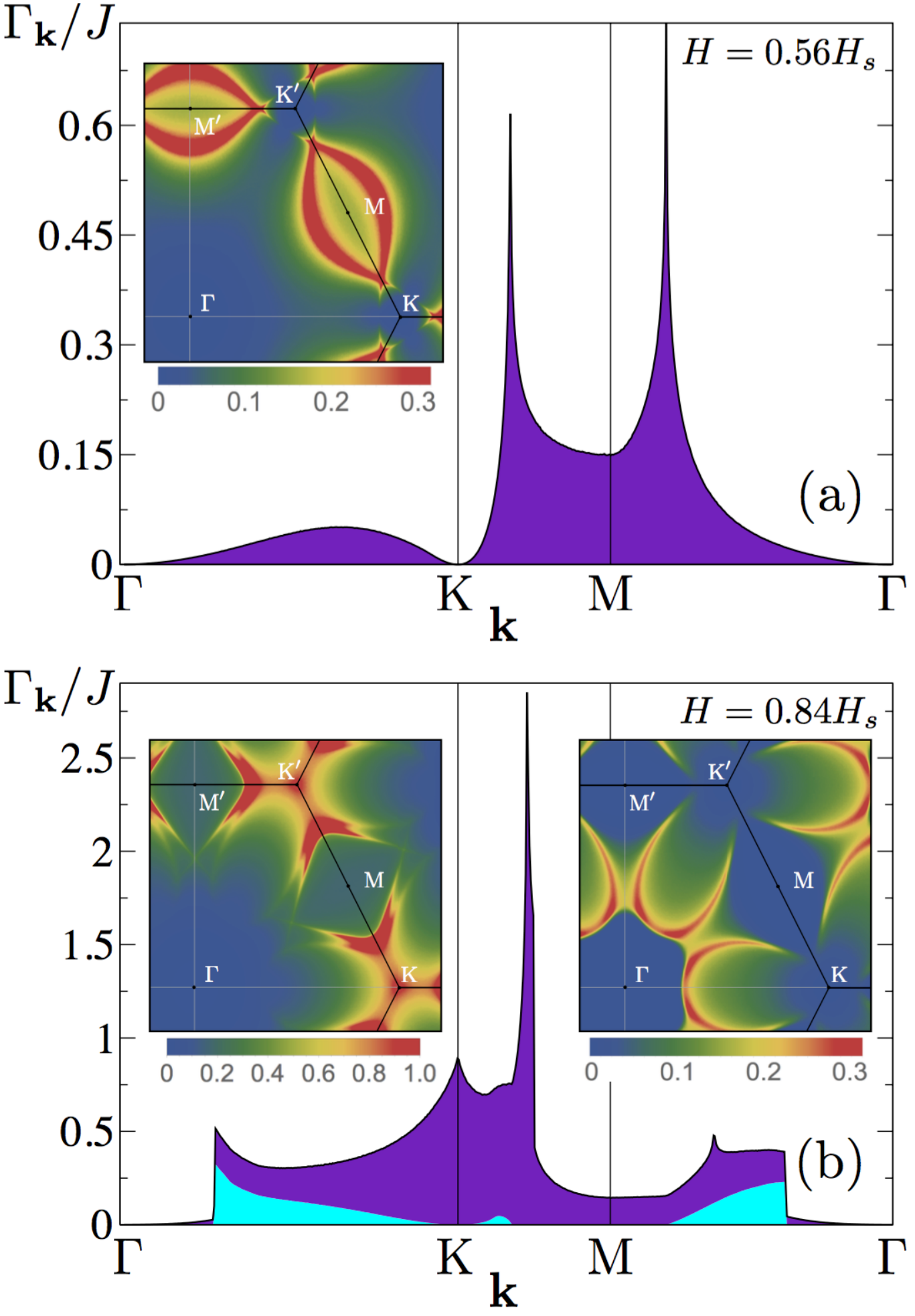}
\caption{(Color online)
\ The field-evolution of the decay rate in the optical branch of magnons for $\Delta=0.5$. Main panels
show $\Gamma_{{\bf k}}$ vs ${\bf k}$ 
along the  $\Gamma\!\rightarrow\!{\rm K}\!\rightarrow\!{\rm M}\!\rightarrow\!\Gamma$ path
with contributions of different channels shown by different shadings [$\Gamma^{2 \rightarrow \{ 1, 1 \}}_{{\bf k}}$ and 
$\Gamma^{2 \rightarrow \{ 2, 1 \}}_{{\bf k}}$ are upper and lower in (b)],  
and insets are the intensity maps for individual channels.
(a) $H\!=\!0.56H_s$ and (b) $H\!=\!0.84H_s$. Field in (a) is below $H^{*}_{2\rightarrow\{2,1\}}$ so that
only $2\rightarrow \{1,1\}$ channel is  active. 
In (b), left and right insets are for $2\rightarrow \{1,1\}$ and $2\rightarrow \{2,1\}$ channels, respectively.
Note different scales in the intensity plots. 
}
\label{fig_gamma2d05}
\end{figure}

In Fig.~\ref{fig_gamma2d0t1} we present an example of the decay rate profiles obtained 
using \eqref{eq_gamma11} and \eqref{eq_gamma122}
along the $\Gamma\!\rightarrow\!{\rm K}\!\rightarrow\!{\rm M}\!\rightarrow\!\Gamma$ cut 
in the ${\bf k}$-space and as the 2D intensity maps, similar to the presentation in Fig.~\ref{fig_gamma1}. The
representative values of $H\!=\!0.84H_s$ and $\Delta\!=\!0$ are chosen so that both channels of the optical mode 
decay are active. In Fig.~\ref{fig_gamma2d0t1}(a), the total decay rate, i.e. the sum of the contributions of the two channels,
is shown and the contributions of individual channels are indicated by different shadings, with the lower shaded 
area corresponding to the  $2\rightarrow \{2,1\}$  channel and the upper one to the  contribution 
of the $2\rightarrow \{1,1\}$ channel. In Figs.~\ref{fig_gamma2d0t1}(b) and (c), the individual 2D intensity
maps of the $\Gamma^{2 \rightarrow \{ 1, 1 \}}_{{\bf k}}$ and $\Gamma^{2 \rightarrow \{ 2, 1 \}}_{{\bf k}}$ are shown 
across the Brillouin zone. The three panels of Fig.~\ref{fig_gamma2d0t1} demonstrate that 
the two decay channels typically dominate different regions of the ${\bf k}$-space, with the $2\rightarrow \{1,1\}$ 
decays most pronounced in the wider area around Brillouin zone boundary, where it can reach 
substantial values in excess of $\Gamma_{{\bf k}} \agt 0.5J$ depending on the field regime, 
see also  Fig.~\ref{fig_gamma2d05}.

% ==============================================================================
% ==============================================================================

Similarly to the acoustic branch decays, the decay rates in  Figs.~\ref{fig_gamma2d0t1} and \ref{fig_gamma2d05}
exhibit a number of singularities, which, upon a closer look, can  be affiliated with the threshold contours discussed in
Sec.~\ref{sec_upbranchkin}. Consider first the $2\rightarrow \{2,1\}$ channel, the lower shaded area in 
Fig.~\ref{fig_gamma2d0t1}(a) and intensity plot in Fig.~\ref{fig_gamma2d0t1}(c). The jump-like 
singularities at $k_1$ and $k_8$ are associated with the crossings of the decay boundary near the 
$\Gamma$ point, see Fig.~\ref{fig_d122_regions1}, which correspond to a generic type of decays.
The boundary at $k_4$ is due to emission of the Goldstone mode, see Sec.~\ref{sec_upbranchkin} and
Fig.~\ref{fig_d122_regions1}, and is clearly not associated with any significant singularity, in agreement with earlier
studies.\cite{triangle} The three other points, $k_2$, $k_6$, and $k_7$, are all associated with 
the same threshold contour, to the decays with an emission of a Dirac-like magnon at the ${\rm K}$ point 
[dashed line in Fig.~\ref{fig_gamma2d0t1}(c)].
However, the $k_2$ point is at the true decay region boundary, while $k_6$ and $k_7$ are inside the decay region
and thus correspond to the saddle point in the two-magnon continuum, hence the logarithmic singularities
for the latter. 

The same $k_6$ and $k_7$ points in Fig.~\ref{fig_gamma2d0t1}(b) are associated with an 
identical contour for  the $2\rightarrow \{1,1\}$ channel, but serve as the decay region boundaries, see 
also Fig.~\ref{fig_d112_regions} for guidance. The $k_2$ point for the $2\rightarrow \{1,1\}$ decay channel
is, in turn, associated with the saddle point in the continuum and thus a logarithmic singularity, but it
nearly coexists with another logarithmic singularity at $k_3$, which, together with the weaker singularity
at $k_5$, is associated with the threshold of decays into two identical magnons [dotted contours in
Fig.~\ref{fig_gamma2d0t1}(b), see also Fig.~\ref{fig_d112_regions}].
Lastly, the kink at the ${\rm K}$ point in Figs.~\ref{fig_gamma2d0t1}(a) and
\ref{fig_gamma2d05}(b) is related to the cone-like structure
in the energy spectrum at this point, see Fig.~\ref{fig_spectrum}, and is not associated with any singularity.
The provided  detailed analysis of singularities demonstrates, once again, 
the useful insights offered by the kinematic consideration 
of Sec.~\ref{sec_kinematics}, as it allows to identify contours of singularities and study
their intricate transformations throughout the Brillouin zone.

While the asymptotic behavior of the decay rates at the $\Gamma$ and ${\rm K}$ points is discussed below in 
some detail, here we highlight a few of its features. First, although it is virtually impossible to see on the scale of 
Fig.~\ref{fig_gamma2d0t1}(a), the decay rate in the $2\rightarrow \{1,1\}$ channel is non-zero in the 
vicinity of the $\Gamma$ point and obeys $\Gamma_{{\bf k}} \propto |{\bf k}|^2$ law, 
see also Fig.~\ref{fig_gamma2d05} for a better representation of this regime and
Fig.~\ref{fig_asymp1}(b) for the asymptotic analysis. As is also
clear from the kinematic consideration in Sec.~\ref{sec_upbranchkin}, the decays in the $2\rightarrow \{2,1\}$
channel are not allowed near the $\Gamma$ point for $|{\bf k}| \lesssim k_1$ in Fig.~\ref{fig_gamma2d0t1}(a). 
One can  observe in the same Figure that the decay rate in the $2\rightarrow \{2,1\}$ channel also
vanishes at the ${\rm K}$ point.
The following analysis demonstrates that it obeys a similar law $\Gamma_{{\bf k}} \propto |{\bf k}-{\bf K}|^2$,
albeit for different reason, see Fig.~\ref{fig_asymp2}(b), while the $2\rightarrow \{1,1\}$ channel yields a finite 
$\Gamma^{2 \rightarrow \{ 1, 1 \}}_{{\bf K}}$ at this field. As one can see in Fig.~\ref{fig_gamma2d05}, 
$\Gamma^{2 \rightarrow \{ 1, 1 \}}_{{\bf k}}$ experiences a change of its asymptotic behavior at
the ${\rm K}$ point vs field from the  $\propto |{\bf k}-{\bf K}|^2$ behavior for the lower fields, same 
as in the $2\rightarrow \{2,1\}$ channel, to the constant value at higher fields. The transition 
happen as the decay boundaries in the $1\rightarrow \{1,1\}$ channel pass through the ${\rm K}$
point, see Fig.~\ref{fig_d1_region}, and allow for the decays of the ${\bf K}$-magnon to the lower states.
This happens at $\lambda\approx 0.69$, which, for the choice of $\Delta=0.5$ in Fig.~\ref{fig_gamma2d05}, 
corresponds to the field $H\approx 0.76H_s$.

Fig.~\ref{fig_gamma2d05} offers some additional details on the field evolution of decays in the 
optical branch of magnons that are highlighted for a representative  $\Delta=0.5$. 
As is mentioned above, the effect of decays in the $2\rightarrow \{1,1\}$ channel is typically 
maximal in the wide area around Brillouin zone boundary, while decays in the 
$2\rightarrow \{2,1\}$ channel, at the fields above the corresponding threshold field 
$H^{*}_{2\rightarrow\{2,1\}}$, are typically clustered along singularity lines in a broad area towards the 
zone center, see Fig.~\ref{fig_gamma2d0t1}(c) and the right inset in Fig.~\ref{fig_gamma2d05}(b). 
Such characteristic features do not only represent rather spectacular shapes, but  can also serve as  important
fingerprints for an experimental identification of the decay-induced phenomena in the spectra
of quantum magnets. Another important note concerns the magnitude of the spectrum broadening 
due to magnon decays. Even with   singularities regularized by the higher-order processes, 
as discussed above,\cite{Mourigal10,99,triangle,RMP} the values of magnon decay rate in 
Figs.~\ref{fig_gamma2d05}(b)   and \ref{fig_decay1e2}
reach very substantial values, $\Gamma_{{\bf k}} \sim J$, possibly as a consequence of 
enhanced quantum fluctuations in the honeycomb-lattice antiferromagnets.

% ==============================================================================
\subsection{Long-wavelength decays}
\label{sec_approx}
% ==============================================================================

The long-wavelength asymptotic behavior of magnon damping provides a useful characterization
of the decay phenomena. It is expected to exhibit universal ${\bf k}$-dependencies and is  
well within the perturbative regime, allowing for explicit analytical evaluations. In the considered 
case of the honeycomb lattice, the vicinities of the 
two high-symmetry points, $\Gamma$ and ${\rm K}$, are of interest.

% ==============================================================================
\subsubsection{${\bf k}\rightarrow\Gamma$}
% ==============================================================================
% ==============================================================================
\begin{figure}[t]
\includegraphics[width=0.99\linewidth]{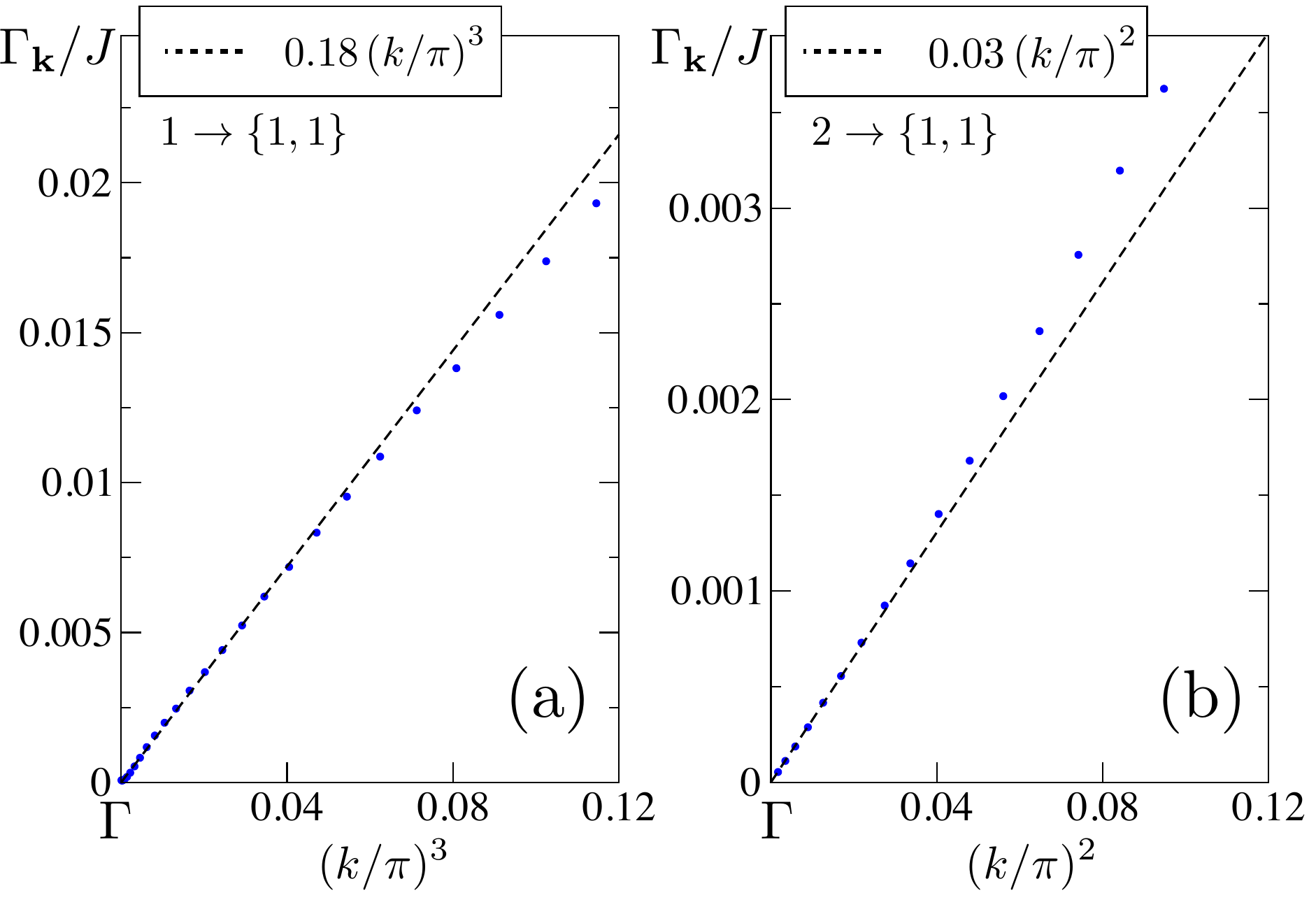}
\caption{(Color online) \ The decay rates in the vicinity of $\Gamma$ point for (a) $1\!\rightarrow \!\{1,1\}$ and 
(b) $2\!\rightarrow\! \{1,1\}$ decay channels for a representative set of $\Delta\!=\!0.5$ and $H\!=\!0.84H_s$. 
Horizontal axes are chosen to emphasize the asymptotic power laws.
Dots are results of numerical integration in (\ref{eq_gamma11})   
and dashed lines are the fits with the corresponding
power laws. }
\label{fig_asymp1}
\end{figure}
% ==============================================================================

The decays of the acoustic branch, $1\rightarrow \{1,1\}$ channel, occur in the proximity of the Brillouin zone center 
only  above $H^{*}_{1\rightarrow\{1,1\}}$ when magnon dispersion has a positive curvature, 
$\varepsilon_{1{\bf k}}\!=\! c|{\bf k}|\!+\!\alpha|{\bf k}|^3$, $\alpha\!>\!0$. This scenario 
is common to  Bose gases, phonon part of the spectrum of $^4$He,\cite{pitaevskii} and some other non-collinear magnets
considered previously.\cite{RMP,Kreisel,99} 
Kinematically, in this common setting,  decays of the long-wavelength excitations are into a pair of other
excitations that belong to an elongated decay contour along the initial momentum, see Ref.~\onlinecite{RMP}. 
Combined with a standard ``hydrodynamic" scaling form of the decay vertex, 
$\Phi^{111}_{{\bf q},{\bf k-q};{\bf k}} \propto\sqrt{|{\bf q}||{\bf k-q}||{\bf k}|}$, a straightforward algebra
leads to the resultant decay rate in 2D $\Gamma^{1\rightarrow \{1,1\}}_{|{\bf k}|\rightarrow 0} \propto |{\bf k}|^3$,
which is identical to the asymptotic behavior of the decay rate in the square-lattice 
antiferromagnets in a field.\cite{99,Mourigal10}
This result is illustrated in Fig.~\ref{fig_asymp1}(a) where we show the decay rate calculated 
numerically using (\ref{eq_gamma11}) 
for a representative set of parameters together with the power-law fit. The  horizontal axis is chosen as 
$\propto k^3$ to make the power-law dependence explicit. One can see that the asymptotic
result works very well up to $|{\bf k}| \approx 0.4 \pi$.

The long-wavelength behavior of the optical mode decays is more intriguing. Formally, according to our 
kinematic considerations in Sec.~\ref{sec_upbranchkin}, 
there is always a finite two-magnon density of states
in the acoustic branch for the optical magnon to decay into using the $2\rightarrow \{1,1\}$ channel, 
even at the $\Gamma$ point itself. Since none of the energies of  excitations involved 
in a decay process of the gapped mode are necessarily small, 
one would naively expect the corresponding decay rate to be finite:
$\Gamma^{2\rightarrow \{1,1\}}_{|{\bf k}|\rightarrow 0} \propto const$. However, one can show that the 
corresponding decay vertex vanishes at $|{\bf k}|\rightarrow 0$ due to the phase factors (\ref{psikqp}), 
yielding instead
\begin{equation}
\Gamma^{2\rightarrow \{1,1\}}_{|{\bf k}|\rightarrow 0} \propto  
\sin^2 \left( \frac{\varphi_{\bf k}+\varphi_{\bf q}+\varphi_{\bf k-q}}{2}\right) \propto |{\bf k}|^2, 
\label{asym211}
\end{equation}
A parallel can be drawn with the square-lattice case,\cite{99,RMP} where the gapped mode 
at ${\bf k}=0$ is  associated with the uniform precession mode and thus can be
rationalized as being protected from decays. In the current case, no direct association 
with the precession mode exists, yet the vertex vanishes at  $|{\bf k}|\rightarrow 0$, 
implying a selection rule. The decay at ${\bf k}=0$ is forbidden instead
because the upper mode is comprised of an antisymmetric combination of the original spin-flips (\ref{eq_phaseshift}),
and should not be able to decay into a symmetric combination of the lower-branch modes.
 
Fig.~\ref{fig_asymp1}(b) demonstrates the validity of the asymptotic behavior in (\ref{asym211})
with the decay rate calculated numerically using (\ref{eq_gamma11})  and its parabolic fit,
the horizontal axis is $\propto k^2$ for clarity. 
Approximation can be seen as valid for $|{\bf k}| \lesssim 0.25 \pi$.

% ==============================================================================
\subsubsection{${\bf k}\rightarrow{\rm K}$}
% ==============================================================================
% ==============================================================================
\begin{figure}[t]
\includegraphics[width=0.99\linewidth]{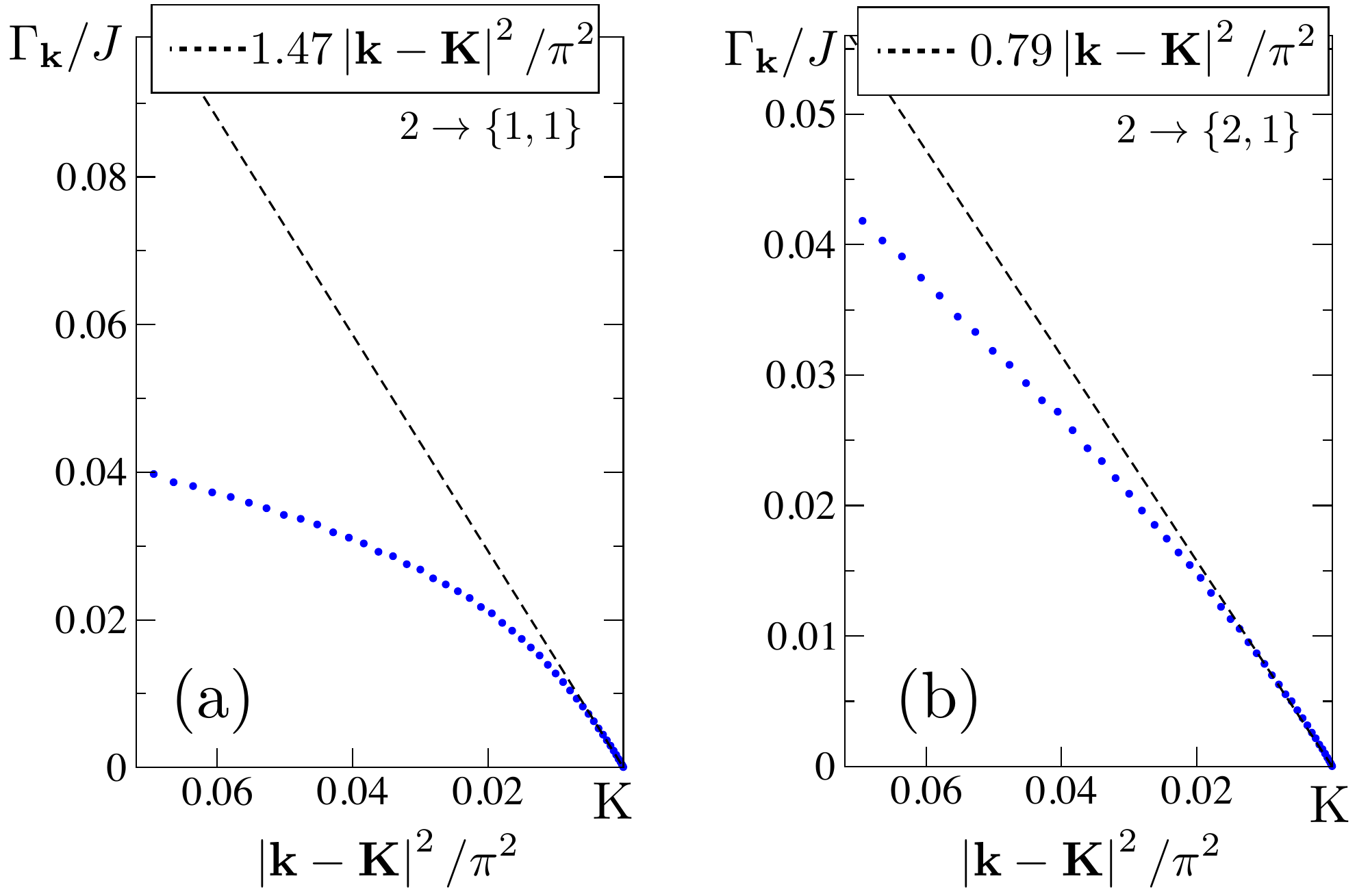}
\caption{(Color online)\ Same as Fig.~\ref{fig_asymp1} for the vicinity of K point.  
(a)  $2\!\rightarrow\! \{1,1\}$ decay channel for $\Delta\!=\!0$, $H\!=\!0.56H_s$, and  (b) 
 $2\!\rightarrow \!\{2,1\}$  channel for $\Delta\!=\!0.5$, $H\!=\!0.84H_s$. }
\label{fig_asymp2}
\end{figure}
% ==============================================================================

As one can see from  representative plots in Fig.~\ref{fig_spectrum}, since 
the two branches are degenerate at the ${\rm K}$ point,  then in its vicinity there is always a possibility 
for an optical magnon to emit a ${\bf q}\approx 0$ 
acoustic magnon and drop onto the lower branch, a decay within the $2\rightarrow \{1,1\}$ channel.
Below the field that correspond 
to  $\lambda\approx 0.69$ via the relation in (\ref{lambda}),  it is the only type of the decay
near the  ${\rm K}$ point which is kinematically allowed. For larger fields, $\lambda\agt 0.69$, the curvature 
of the acoustic branch is bent up significantly so that the decays into two magnons with half-momenta of the
${\rm K}$ point become possible. This transition corresponds to the ``frontline'' of the decay region in 
Fig.~\ref{fig_d1_region} passing through the ${\rm K}$ point. 
Thus, in larger fields, the  momenta involved in a typical decay from the ${\rm K}$ point are not small 
and   lead to a finite decay rate in  the $2\rightarrow \{1,1\}$ channel, a transition 
highlighted in the discussion of Fig.~\ref{fig_gamma2d05} above.

At yet somewhat higher fields, which correspond to $\lambda\!=\!\sqrt{3}\!-\!1\!\approx\!0.73$,
the velocity of the Dirac-like excitation at the ${\rm K}$ point
exceeds the sound velocity of the Goldstone mode, making the decay in the $2\!\rightarrow\! \{2,1\}$ channel
possible, in which a decaying magnon from the optical branch remains in the same branch but sheds a, now slower, 
${\bf q}\!\rightarrow \!0$ magnon. Curiously, this corresponds to the fields somewhat higher  than the threshold field 
$H^{*}_{2\rightarrow\{2,1\}}$  discussed in Sec.~\ref{sec_upbranchkin},  as the current 
consideration requires   decay regions in Fig.~\ref{fig_d122_regions2} to extend to the ${\rm K}$ point.

In both cases, $2\rightarrow \{1,1\}$ channel for $\lambda< 0.69$ and $2\rightarrow \{2,1\}$ channel for
$\lambda>0.73$, the asymptotic behavior near the ${\rm K}$ point is related to the 
emission of the Goldstone ${\bf q}\rightarrow 0$ magnon. Because of that, the 
corresponding decay vertices carry the same smallness in ${\bf q}$
\begin{equation}
\Phi^{112}_{{\bf q},{\bf k-q}; {\bf k}} \propto \Phi^{122}_{{\bf q},{\bf k-q}; {\bf k}} \propto \sqrt{|{\bf q}|}.
\end{equation}
In either of the cases, there is no strong restriction on the angle and thus the 2D density of states also contributes
one power of ${\bf k}-{\bf K}$ to the resultant decay rate, yielding
\begin{equation}
\Gamma^{2\rightarrow \{1,1\}}_{{\bf k}} \propto\Gamma^{2\rightarrow \{2,1\}}_{{\bf k}} \propto 
|{\bf k}-{\bf K}|^2.
\end{equation}
This asymptotic behavior is illustrated in Figs.~\ref{fig_asymp2}(a) and (b), 
with the decay rates in the two channels calculated  using (\ref{eq_gamma11}) 
and (\ref{eq_gamma122}), respectively, and compared to parabolic fits,
with the horizontal axis proportional to $|{\bf k}-{\bf K}|^2$ for clarity. 
This asymptotic  approximation for the $2\rightarrow \{1,1\}$ channel is valid 
in a somewhat narrower range, $|{\bf k}-{\bf K}| \lesssim 0.1 \pi$, because of significant nonlinearities in the upper
branch spectrum.

% ==============================================================================
\section{Dynamical structure factor}
% ==============================================================================
\label{sec_strfac}
% ==============================================================================

Using  decay rates discussed above, in this Section we calculate the dynamical structure factor 
$\mathcal{S} ({\bf q},\omega)$ and magnon spectral function $A({\bf q},\omega)$ in the leading $1/S$ order.
We show that magnon decay processes lead to a broadening of quasiparticle peaks and to a redistribution of 
spectral weight. These results should be directly relevant to the prospective neutron scattering experiments in the
honeycomb-lattice materials in external field.

Previously, a detailed consideration of the structure factor in noncollinear antiferromagnets was given for the cases 
of the square-lattice antiferromagnet in a field\cite{Mourigal10,Fuhrman13} and for the triangular-lattice
antiferromagnet.\cite{triSqw} More recently, the leading-order, $1/S$ analysis of the structure factor
was provided for the anisotropic kagome-lattice models.\cite{kagome15} 
Our current consideration is similar to the square-lattice case,\cite{Fuhrman13} but the main  
difference is in the non-Bravais nature of the honeycomb lattice, which makes it close to the 
kagome-lattice case in several aspects.\cite{kagome15}
Similarly to the latter,  we restrict ourselves to the leading $1/S$ order of the spin-wave theory, thus 
implying  applicability of our results to $S\agt 1$ systems.

% ==============================================================================
\subsection{$\mathcal{S} ({\bf q},\omega)$ in $1/S$, kinematic formfactors}
% ==============================================================================

We begin with the structure factor for the the non-Bravais lattices, which 
corresponds to  the correlation function of the unit cell magnetizations\cite{Kopietz}
\begin{equation}
\mathcal{S}^{\alpha_0 \beta_0} ({\bf q},\omega) =-\frac{1}{\pi} \text{Im} \int_{-\infty}^{\infty} \, dt \, e^{i\omega t}
\, {\cal G}^{\alpha_0 \beta_0}({\bf q},t)\, ,
\label{Sqw0}
\end{equation}
where the time-ordered, $T=0$ spin Green's function is  
\begin{equation}
{\cal G}^{\alpha_0 \beta_0}({\bf q},t)=-i\left\langle {\cal T} M^{\alpha_0}_{{\bf q}}(t) M^{\beta_0}_{-{\bf q}}(0) \right\rangle,
\label{Sqw1}
\end{equation}
and the Fourier transform of the $\alpha_0$-component of the magnetization 
\begin{equation}
M^{\alpha_0}_{{\bf q}}=\sum_{i,\alpha} S^{\alpha_0}_{i,\alpha} \, e^{i{\bf q} ({\bf r}_i+{\bm \rho}_\alpha)} \, ,
\label{Mq1}
\end{equation}
involves the sum of the spin components in a unit cell, where $\alpha$ numerates the atoms 
and ${\bm \rho}_\alpha$ are the coordination vectors, see \eqref{eq_fourier}. The $\alpha_0$ and $\beta_0$ indices
correspond to the laboratory reference frame $\{x_0,y_0,z_0\}$ in Fig.~\ref{fig_lattice}.

Since the inelastic neutron-scattering cross section is proportional to a linear combination of the \emph{diagonal} 
components of the correlation function in (\ref{Sqw0}) and since the spins in our consideration
form a coplanar structure in the $(x_0,z_0)$ plane, it is convenient to separate the structure factor
into the in-plane and the out-of-plane parts
\begin{equation}
\mathcal{S}^{\parallel}=\mathcal{S}^{x_0 x_0}+\mathcal{S}^{z_0 z_0},\quad\mathcal{S}^{\perp}=\mathcal{S}^{y_0 y_0}\, ,
\label{eq_inoutSqw}
\end{equation}
where we have also assumed equal contribution of all three components.

Let us consider $\mathcal{S}^{z_0 z_0}$ as an example. Using (\ref{eq_localtransform}) and 
(\ref{HP}), in the leading $1/S$ order 
\begin{equation}
M^{z_0}_{{\bf q}}(t)\approx\cos\theta \sqrt{\frac{S}{2}} 
\sum_{\alpha} \left( a_{\alpha {\bf q}}(t) +a^{\dagger}_{\alpha -{\bf q}}(t) \right)  \, ,
\label{Mq2}
\end{equation}
which, after straightforward algebra  with the unitary  
\eqref{eq_phaseshift} and Bogolyubov \eqref{eq_bogolyubov} transformations, yields 
\begin{eqnarray}
\mathcal{S}^{z_0 z_0}({\bf q},\omega)=\frac{S}{4}\cos^2 \theta \sum_\mu A_\mu ({\bf q},\omega) \, 
{\widetilde{\cal F}}_{\mu{\bf q}}^{z_0 z_0}
\label{Sz0z0}
\end{eqnarray}
where we have introduced  magnon spectral function 
$A_\mu ({\bf q},\omega)=-\frac{1}{\pi} \text{Im}  G_{\mu} ({\bf q},\omega)$
of the $\mu=1,2$ branch and the auxiliary function is
\begin{eqnarray}
{\widetilde{\cal F}}_{\mu{\bf q}}^{z_0 z_0}=\left(u_{\mu{\bf q}}+v_{\mu{\bf q}} \right)^2
\sum_{\alpha\alpha'} V^{\mu \alpha} V^{\mu \alpha'}  e^{ \varphi_{{\bf q}}\sigma^y_{\alpha\alpha'}}\, ,
\label{Fz0z0}
\end{eqnarray}
with $\sigma^y$ being the Pauli matrix, see Sec.~\ref{sec_swt} for the rest of the notations.
Using the simplicity of the matrix ${\bf \hat{V}}$ in (\ref{V}), Eq.~(\ref{Fz0z0}) can be simplified, finally leading to
\begin{eqnarray}
&&\mathcal{S}^{z_0 z_0}({\bf q},\omega)=S \cos^2 \theta  \left[  
\left(u_{1{\bf q}}+v_{1{\bf q}} \right)^2 \cos^2 \frac{\varphi_{\bf q}}{2} A_1 ({\bf q},\omega) \right.\nonumber\\
&&\phantom{\mathcal{S}^{z_0 z_0}({\bf q},\omega)=}
+\left. \left(u_{2{\bf q}}+v_{2{\bf q}} \right)^2 \sin^2 \frac{\varphi_{\bf q}}{2} A_2 ({\bf q},\omega) \right].
\end{eqnarray}
Calculating  $\mathcal{S}^{x_0 x_0}$ and $\mathcal{S}^{y_0 y_0}$ in a similar fashion gives
the in-plane and out-of-plane structure factors
\begin{eqnarray}
&&\mathcal{S}^{\parallel}({\bf q},\omega)=S\left({\cal F}^{\parallel}_{1{\bf q}}A_1({\bf q},\omega)
+{\cal F}^{\parallel}_{2{\bf q}}A_2({\bf q},\omega)\right),
\label{Spar}\\
&&\mathcal{S}^{\perp}({\bf q},\omega)=S\Big({\cal F}^{\perp}_{1{\bf q}}A_1({\bf q},\omega)
+{\cal F}^{\perp}_{2{\bf q}}A_2({\bf q},\omega)\Big),
\label{Sperp}
\end{eqnarray}
where we introduced kinematic  formfactors
\begin{eqnarray}
&&{\cal F}^{\parallel}_{1{\bf q}}= \left( \sin^2\theta\sin^2 \frac{\varphi_{\bf q}}{2}  + 
\cos^2\theta\cos^2 \frac{\varphi_{\bf q}}{2}\right) \left(u_{1{\bf q}}+v_{1{\bf q}} \right)^2,  \nonumber\\
&&{\cal F}^{\parallel}_{2{\bf q}}=\left( \sin^2\theta\cos^2 \frac{\varphi_{\bf q}}{2}  + 
\cos^2\theta \sin^2 \frac{\varphi_{\bf q}}{2}\right) \left(u_{2{\bf q}}+v_{2{\bf q}} \right)^2,  \nonumber\\
&&{\cal F}^{\perp}_{1{\bf q}}=\sin^2 \frac{\varphi_{\bf q}}{2}\,  \left(u_{1{\bf q}}-v_{1{\bf q}} \right)^2,  \\
&&{\cal F}^{\perp}_{1{\bf q}}=\cos^2 \frac{\varphi_{\bf q}}{2} \, \left(u_{2{\bf q}}-v_{2{\bf q}} \right)^2  \, .\nonumber
\label{eq_Finout}
\end{eqnarray}

% ==============================================================================
\begin{figure}[t]
\includegraphics[width=0.99\linewidth]{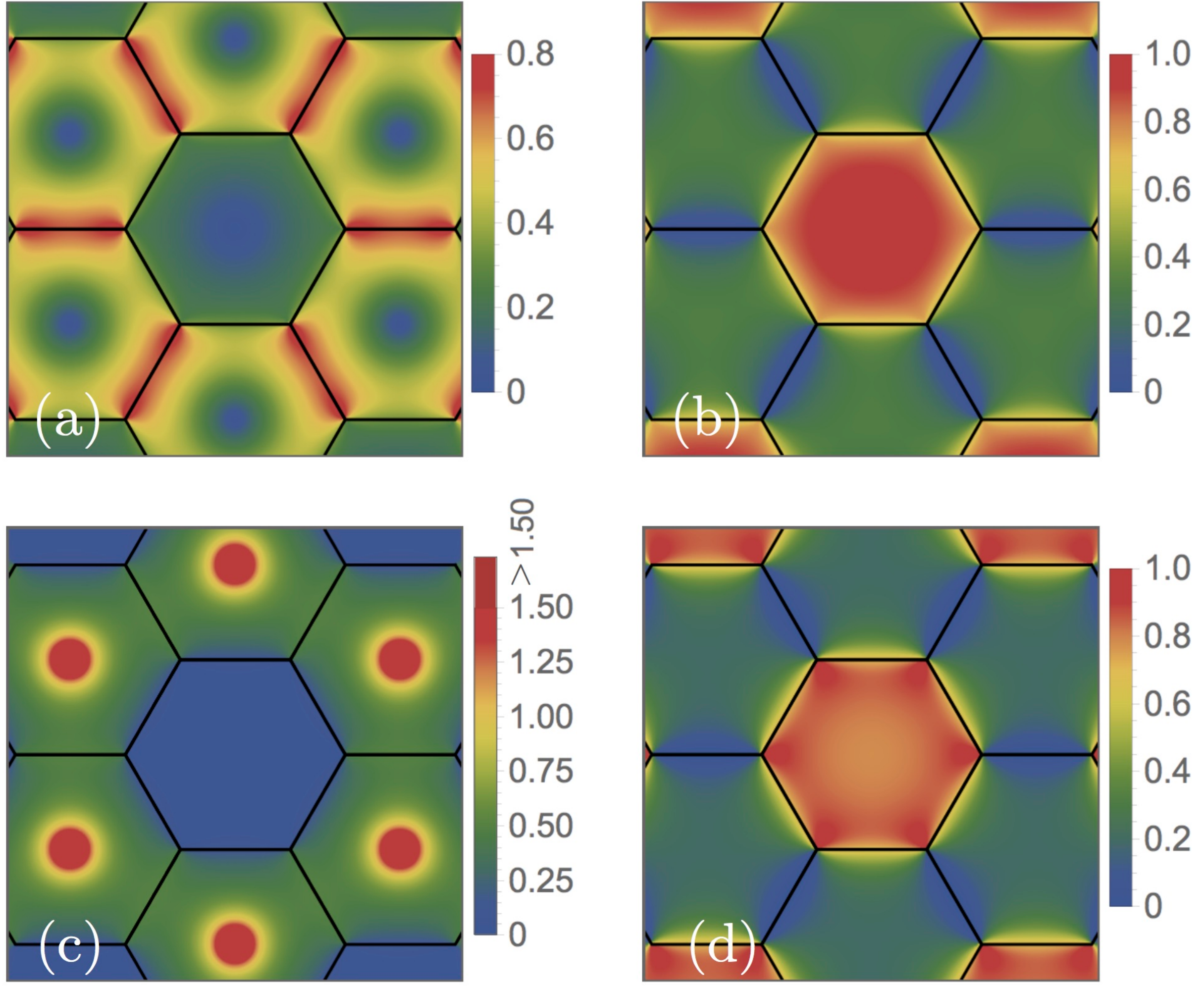}
\caption{(Color online)\ The 
kinematic formfactors for $\Delta\!=\!0$, $H\!=\!0.48H_s$ across several Brillouin zones. 
The formfactors 
for the in-plane $\mathcal{S}^{\parallel}({\bf q},\omega)$: (a)  lower branch  ${\cal F}^{\parallel}_{1{\bf q}}$, 
(b)  upper branch ${\cal F}^{\parallel}_{2{\bf q}}$.  The formfactors 
for the out-of-plane $\mathcal{S}^{\perp}({\bf q},\omega)$: (c)  lower branch  ${\cal F}^{\perp}_{1{\bf q}}$, 
(d) upper branch  ${\cal F}^{\perp}_{2{\bf q}}$.}
\label{fig_sfmasks}
\end{figure}
% ==============================================================================

As in the case of the more standard Bravais-lattice considerations, the kinematic formfactors
serve as modulators of intensity of the magnon spectral functions in ${\bf q}$-space. 
The difference here is in the ${\bf q}$-modulation between \emph{different} Brillouin zones, as in the 
non-Bravais lattices the kinematic formfactors are typically suppressed in one while are maximal in the other
Brillouin zones, the property demonstrated in Fig.~\ref{fig_sfmasks} for representative anisotropy and field values. 
This effect is similar in spirit to the Bragg peak extinction for the elastic scattering in non-Bravais lattices 
and was recently discussed for the kagome-lattice models.\cite{kagome15} 
Because of this ${\bf q}$-modulation,  prospective neutron-scattering experiments should be able to 
separate contributions of different excitation branches by selecting component of the 
structure factor in a specific Brillouin zone.

% ==============================================================================
\subsection{Spectral functions $A_\mu({\bf q},\omega)$ and $\mathcal{S} ({\bf q},\omega)$}
% ==============================================================================

% ==============================================================================
\begin{figure*}
\includegraphics[width=0.99\textwidth]{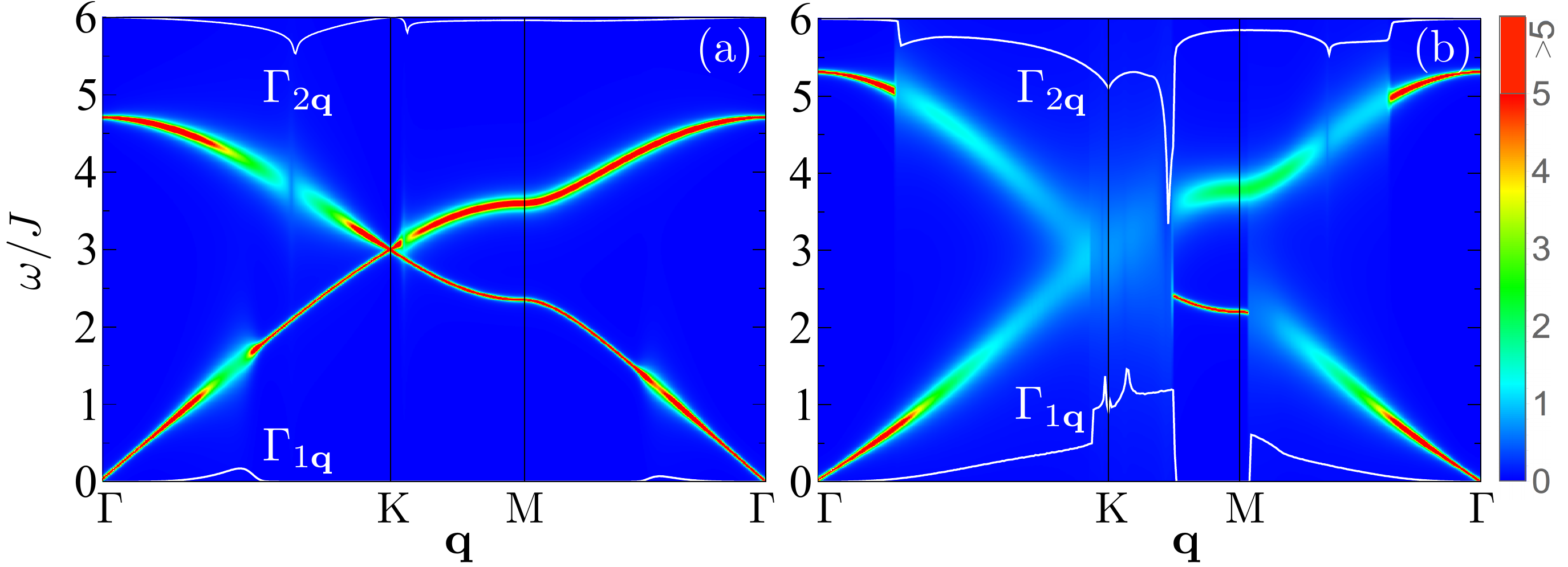}
\caption{(Color online) \ The intensity map of the spectral functions $A_\mu({\bf q},\omega)$ in units of $J^{-1}$ 
from the Green's function in \eqref{eq_GF} for $S\!=\!1$ and (a) $\Delta\!=\!0$ and $H\!=\!0.48H_s$, (b) 
$\Delta\!=\!0.5$ and $H\!=\!0.84H_s$. A small value of  $\delta\!=\!2.5\cdot 10^{-3}J$ 
was added to $\Gamma_{{\bf q}}$.
Broadening of the lower and upper branches, $\Gamma_{1{\bf q}}$ and $\Gamma_{2{\bf q}}$, 
are shown by solid lines in the lower and upper parts of the graphs, 
respectively. Upper cutoff of intensity is $5/J$, which corresponds to a minimum broadening $0.064J$. See text for   
discussion.}
\label{fig_spectfunc}
\end{figure*}
% ==============================================================================

In the non-interacting limit, or at $S\gg1$, the spectral function is a $\delta$-function at 
$\omega=\varepsilon_{\mu{\bf q}}$ for any given ${\bf q}$. 
To demonstrate the effects induced by magnon interaction, we use
a simplified form of the Green's function 
\begin{equation}
G^{-1}_{\mu} ({\bf q},\omega)=\omega-\varepsilon_{\mu{\bf q}}+i\Gamma_{\mu{\bf q}}\, ,
\label{eq_GF}
\end{equation}
with the magnon self-energy taken on-shell and approximated by its imaginary part,
$\Sigma_\mu({\bf q},\varepsilon_{\mu{\bf q}})\approx -i\Gamma_{\mu{\bf q}}$,
where $\Gamma_{\mu{\bf q}}$ is the decay rate of the $\mu$-magnon  (\ref{eq_gammak}) 
in all possible decay channels. Such a simplification neglects the higher-order $1/S$ contributions, the
$O(S^0)$ contributions of the same order from the so-called quartic terms, and the real part of the self-energy.
However, it keeps the most important \emph{qualitatively} new effect of the cubic anharmonicities in the
spectrum,---broadening due to magnon decays.
Thus, our approach is close in spirit to the one in Ref.~\onlinecite{Mourigal10},  albeit without the 
self-consistency in  the decay rate [called iSCBA scheme in Ref.~\onlinecite{Mourigal10}]. 
It is, therefore, expected to be applicable to the $S\agt 1$ systems, but can also be instructive for the 
$S=1/2$ case. The same approximation has been used recently for the theoretical 
analysis of the spectral properties of the $S=5/2$ kagome-lattice system, Fe-jarosite.\cite{kagome15}

Since the $\omega$-dependence is neglected in the self-energy in (\ref{eq_GF}), the spectral function
$A_\mu({\bf q},\omega)$ takes the form of a simple Lorentzian for any ${\bf q}$, which is obviously a 
simplification too. Nevertheless, the offered consideration is still immensely instructive as our results are expected 
to be quantitative with respect to the variation of the magnitude of the 
broadening effect in the ${\bf q}$-space, identifying regions with stronger and weaker decays, 
and of the spectral weight redistribution. However, our analysis will only be qualitative 
regarding the subtler $\omega$-structures in the spectrum, such as the ones studied in a related problem 
of the spectral properties of the square-lattice antiferromagnet in a field.\cite{Fuhrman13}

Here we offer two representative examples of the spectral functions $A_\mu({\bf q},\omega)$ of both 
magnon bands for a prospective $S\!=\!1$ honeycomb-lattice $XXZ$ antiferromagnet, see Fig.~\ref{fig_spectfunc},
which shows the intensity plots of  $A_\mu({\bf q},\omega)$ in the $\omega-{\bf q}$ plane for  
$\mu\!=\!1$ and $2$ along the usual $\Gamma\!\rightarrow\!{\rm K}\!\rightarrow\!{\rm M}\!\rightarrow\!\Gamma$ 
path in the Brillouin zone. $\Gamma_{1{\bf q}}$ and $\Gamma_{2{\bf q}}$, which correspond to half-width at 
half-maximum of the spectral peaks, are also shown for reference by the solid lines 
in the lower and upper parts of the graphs, respectively.  
The upper cutoff in intensity is  $5/J$, which translates into the minimal line broadening of 
$0.064J$ that should be easily resolvable by the modern neutron-scattering experiments 
for  appropriate values of $J$. 

First choice of parameters, $\Delta\!=\!0$ and  $H\!=\!0.48H_s$, in Fig.~\ref{fig_spectfunc}(a) 
corresponds to the field larger than the threshold field in the $1\!\rightarrow\!\{1,1\}$ decay channel, i.e. 
above $H^{*}_{1\rightarrow\{1,1\}}$ when the lower branch magnon dispersion has a positive curvature, 
but below the threshold in the $2\!\rightarrow\!\{2,1\}$ channel, $H^{*}_{2\rightarrow\{2,1\}}$, 
see Fig.~\ref{fig_phasediagram1} and Sec.~\ref{sec_kinematics}. Thus the decay processes in both magnon 
branches are only into  pairs of  magnons in the lower, acoustic branch, i.e. $\mu\!\rightarrow\!\{1,1\}$.
One can see in  Fig.~\ref{fig_spectfunc}(a) that there is a clear decay boundary in the broadening of the lower branch,
with the acoustic magnons still sharply defined for large ${\bf q}$ toward the Brillouin zone boundary,
see also Fig.~\ref{fig_d1_region}.
At the same time, the quasiparticle peaks in the upper branch are broadened for any ${\bf q}$ except the $\Gamma$ and 
${\rm K}$ points, see Sec.~\ref{sec_cubic}. The overall magnitude of the broadening can be described as moderate, 
$\Gamma_{{\bf q}}\alt 0.2J-0.3J$ for most ${\bf q}$ values,
which is still well within the range of detectability by the modern neutron-scattering experiments.
As is discussed in Sec.~\ref{sec_cubic}, the divergent singularities in 
$\Gamma_{\mu{\bf q}}$, which result into vanishing spectral weight at select ${\bf q}$ points in Fig.~\ref{fig_spectfunc}(a),
are going to be regularized by the higher-order decay processes.

The effects of decays are much more dramatic in Fig.~\ref{fig_spectfunc}(b), which corresponds to the 
same choice of $S=1$, but the anisotropy and the field are $\Delta\!=\!0.5$ and  $H\!=\!0.84H_s$.
For this value of $\Delta$, the chosen field is above both of the  threshold fields $H^{*}_{1\rightarrow\{1,1\}}$ and 
$H^{*}_{2\rightarrow\{2,1\}}$, see Fig.~\ref{fig_phasediagram1} and Sec.~\ref{sec_kinematics}. 
Fig.~\ref{fig_spectfunc}(b) shows clear and rather spectacular termination-point-like transitions in both 
acoustic and optical branches. In the latter, the thresholds near the $\Gamma$ point are related to the 
decay region threshold for the $2\!\rightarrow\!\{2,1\}$ channel, see discussion of 
Figs.~\ref{fig_gamma2d0t1} and \ref{fig_gamma2d05} and Fig.~\ref{fig_d122_regions2}.
The difference from the previously studied models\cite{RMP} is that such an abrupt transition can
happen \emph{within} the decay region, i.e. for a part of the magnon spectrum which already has a finite 
lifetime. This is because of the multiple decay channels having different kinematic constraints on the decays.

The intensity plot also shows other features, such as  partial recovery of the upper branch around
the ${\rm M}$ point where decays become more modest and the ``watershed''-like wipeout of both branches
in the proximity of the ${\rm K}$ point. Perhaps the most remarkable feature of the spectrum in Fig.~\ref{fig_spectfunc}(b)
is the perfectly well-defined portion of the lower branch magnon in the stretch of ${\bf q}$ values
around the ${\rm M}$ point, which is surrounded by the regions that appear to be free from any 
quasiparticle-like excitations because of strong overdamping, see also Fig.~\ref{fig_decay1e2}.

Overall, the magnitude of broadening in Fig.~\ref{fig_spectfunc}(b) 
is rather significant, even assuming all the singularities to be regularized, 
reaching  $\Gamma_{{\bf q}}\sim 0.5J-1.0J$ for a wide range of ${\bf q}$ and for both magnon branches.
This leads to strong suppression of the quasiparticle peaks
despite the somewhat elevated value of the spin, $S=1$, and can possibly be related to  
enhanced quantum fluctuations in the honeycomb-lattice antiferromagnets as discussed in Sec.~\ref{sec_cubic}.
Although our consideration is partially qualitative, one can expect that the massive spectral 
weight redistribution and significant suppression of the quasiparticle peaks together with the rather peculiar 
distribution of strongly and weakly affected regions in the ${\bf q}$-space will survive a more rigorous treatment
of the problem.\cite{Lauchli}

% ==============================================================================
\begin{figure*}
\includegraphics[width=0.95\linewidth]{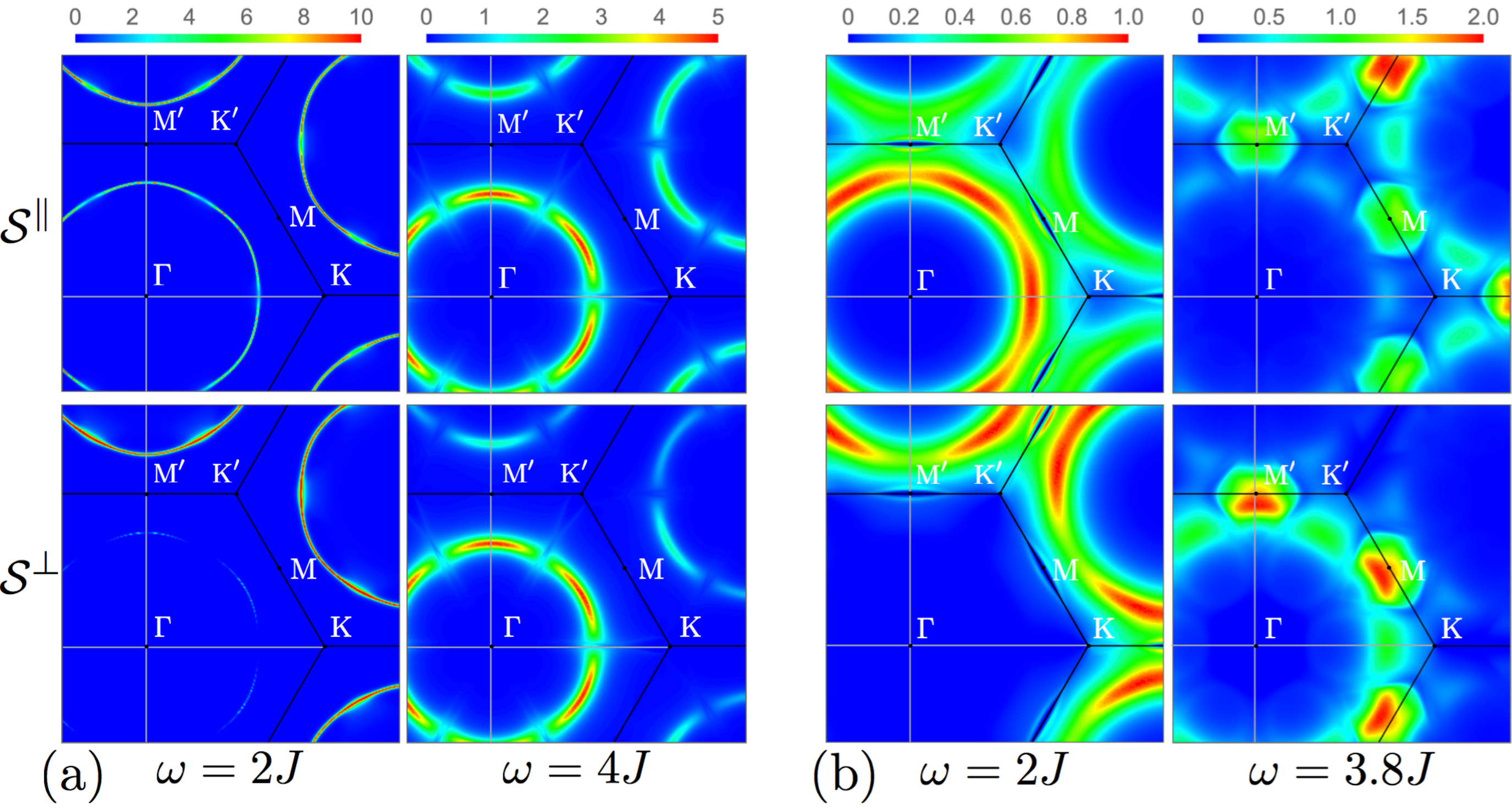}
\caption{(Color online)\ The intensity maps of the constant-energy cuts of the
in-plane and out-of-plane structure factors, 
$\mathcal{S}^{\parallel}({\bf q},\omega)$ and $\mathcal{S}^{\perp}({\bf q},\omega)$, (\ref{Spar}) and (\ref{Sperp}), 
for $S=1$ and (a) $\Delta=0$ and $H=0.48H_s$, (b) $\Delta=0.5$ and $H=0.84H_s$, same as in Fig.~\ref{fig_spectfunc}. 
Values of constant energies are as indicated in the graph. Modulation of intensities between Brillouin zones is due to 
kinematic formactors; see Fig.~\ref{fig_sfmasks} and text.}
\label{fig_strfac}
\end{figure*}
% ==============================================================================

A different view on the unusual spectral features of the $XXZ$ honeycomb-lattice antiferromagnet in a field 
can be gained via the constant-energy cuts of the $\mathcal{S} ({\bf q},\omega)$, 
the approach often used in the modern inelastic neutron-scattering experiments.
In Fig.~\ref{fig_strfac} we show several examples of them for $S=1$ and the same parameters as in 
 Fig.~\ref{fig_spectfunc}.
Fig.~\ref{fig_strfac}  shows separately the intensity maps of the in-plane and the out-of-plane structure-factor 
components,  $\mathcal{S}^{\parallel}({\bf q},\omega)$ and $\mathcal{S}^{\perp}({\bf q},\omega)$
given by (\ref{Spar}) and (\ref{Sperp}), for two representative energies and in the window of the 
${\bf q}$-space, which offers a partial view of several neighboring Brillouin zones.

The set in Fig.~\ref{fig_strfac}(a) demonstrates the $\omega$-cuts for the same $\Delta$ and  $H$ as 
in Fig.~\ref{fig_spectfunc}(a) for $\omega=2J$ and $\omega=4J$. At the lower energy,  
one can observe circular cuts of the cone-like magnon dispersion, modulated in intensity by the 
kinematic formfactors from Fig.~\ref{fig_sfmasks} discussed above,
with little deviation from the picture expected from the sharp, well-defined quasiparticle excitations. 
For the higher energy cut, such deviations become more apparent, with the broadening exhibiting 
an intriguing modulation along the similar circular cuts. This is in accord with Fig.~\ref{fig_spectfunc}(a)
which showed only moderate damping primarily in the optical branch of magnons.

A very strong deviation from sharp quasiparticle contours, characterized by  intense broadening and
massive redistribution of the spectral weight into different regions of the ${\bf q}$-space, 
together with the non-trivial character of such redistribution,
can be observed for both energies in the set in Fig.~\ref{fig_strfac}(b) that uses the same parameters
as in  Fig.~\ref{fig_spectfunc}(b). 
This is also in agreement with the exposition in Fig.~\ref{fig_spectfunc}(b), which
indicated much stronger broadening.
In both cases in Fig.~\ref{fig_strfac}, the intensity is strongly modulated between  Brillouin zones
due to kinematic formfactors. For the lower-energy cuts of the lower branch,
both the in-plane and the out-of-plane structure factors are enhanced in the second and third Brillouin zones,
while for the cut at higher energy, which concerns the optical part of the spectrum, 
the structure factors are enhanced in the first Brillouin zone and are  suppressed in the second and third Brillouin zones.
 
Altogether,  the constant-$\omega$ cuts of $\mathcal{S} ({\bf q},\omega)$ in Fig.~\ref{fig_strfac} together with  
intensity maps of $A_\mu({\bf q},\omega)$ in Fig.~\ref{fig_spectfunc}
provide a distinct and detailed representation of the strong decay-induced quantum effects  
in the dynamical response of the honeycomb-lattice antiferromagnet in a field, which take their origin in the 
anharmonic magnon couplings.
These results are general and should apply to various honeycomb-based systems and
we expect future experiments in applied magnetic field on such materials to confirm our findings. 

% ==============================================================================
\section{Conclusions}
% ==============================================================================

To summarize, we have developed  a formalism of the 
nonlinear spin-wave theory for the nearest-neighbor honeycomb-lattice $XXZ$ model in a field 
and have provided a comprehensive study of the dynamical effects that are induced by the two-magnon decay processes 
in its spectrum at zero temperature.  
We have given a detailed  analysis of important and distinct features that make 
decays of spin excitations in the honeycomb-lattice antiferromagnets significantly richer than in the previously
studied cases, such as non-Bravais nature of the system and the Dirac-like points in the spectrum.

In particular, the decays of the optical mode into two  acoustic ones are allowed for any non-zero field, 
while the other decay channels, such as acoustic into two acoustic or optical
into a mix of optical and acoustic, require  threshold fields to be exceeded. 
A separate degree of complication is added by the Dirac-like degeneracy points connecting 
magnon branches that contribute to the intriguing field-evolution of the decay regions in the ${\bf q}$
space, which has been studied in detail in the present work together with the concomitant singularities originating 
from the coupling to the two-magnon continuum. We have illustrated 
our findings by the maps of the field-dependent decay regions, singularity contours, and decay rates in all the channels
for several representative values of anisotropy and field. 
We have also provided a meticulous analysis of the singularities in decay rates 
and have studied   asymptotic behavior of the latter in the vicinities of the high-symmetry points. 

Lastly,  we have derived analytical expressions for the inelastic neutron-scattering spin-spin structure factor in the 
leading $1/S$ order and have demonstrated its qualitatively distinct features that take their origin in 
the decay-induced magnon dynamics.
In passing, we have shown that the kinematic formfactors can modulate observed signal in different Brillouin zones
owing to the non-Bravais nature of the system, thus offering a clear separation of the contribution from different modes.
 The main effects of the decays are in the clear and dramatic broadening of the quasiparticle peaks and in the
strong spectral weight redistribution. These findings are illustrated by the
intensity maps of the magnon spectral functions and by constant-energy cuts of the dynamical 
structure factor for representative values of the field and anisotropy.

We expect a qualitative applicability of our results to the other related models and
anticipate the future neutron-scattering studies to contribute to a further progress in this area.

%----------------------------------------------------------------------
\begin{acknowledgments}

We are indebted to Michael Zhitomirsky, conversations with whom have initiated this study, for important 
discussions. 
 %We acknowledge useful  discussions with  . 
This work was supported by the U.S. Department of Energy,
Office of Science, Basic Energy Sciences under Award \# DE-FG02-04ER46174.
A.~L.~C. would like to thank  the Kavli Institute for Theoretical Physics where part of this work was done. 
The work at  KITP was supported in part by NSF Grant No. NSF PHY11-25915.

\end{acknowledgments}
%----------------------------------------------------------------------

% ==============================================================================

% ==============================================================================

\end{document}